\documentclass[onecolumn]{aastex62}

\usepackage{amsmath}
\usepackage{threeparttable}
\usepackage{footnote}
\usepackage{color}
\usepackage[figuresright]{rotating}
\usepackage[hang,flushmargin]{footmisc}

\def\simgt{\lower 2pt \hbox{$\, \buildrel {\scriptstyle >}\over {\scriptstyle \sim}\,$}}
\def\simlt{\lower 2pt \hbox{$\, \buildrel {\scriptstyle <}\over {\scriptstyle \sim}\,$}}

\received{}
\revised{}
\accepted{}

\shorttitle{Searching Spectroscopic Binaries with Machine Learning}
\shortauthors{Liu et al.}

\begin{document}

\title{Double-lined Spectroscopic Binaries from the LAMOST Low-Resolution Survey}

\correspondingauthor{Bo Zhang; Jianfeng Wu}
\email{bozhang@nao.cas.cn; wujianfeng@xmu.edu.cn}

\author[0000-0002-7600-1670]{Junhui Liu}
\affiliation{Department of Astronomy, Xiamen University, Xiamen, Fujian 361005, People's Republic of China}
\affiliation{Research School of Astronomy \& Astrophysics, Australian National University, Cotter Rd., Weston, ACT 2611, Australia}

\author[0000-0002-6434-7201]{Bo Zhang}
\affiliation{Key Laboratory of Space Astronomy and Technology, National Astronomical Observatories, Chinese Academy of Sciences, Beijing 100101, People's Republic of China}

\author[0000-0001-7349-4695]{Jianfeng Wu}
\affiliation{Department of Astronomy, Xiamen University, Xiamen, Fujian 361005, People's Republic of China}

\author[0000-0001-5082-9536]{Yuan-Sen Ting}
\affiliation{Department of Astronomy, The Ohio State University, Columbus, OH 43210, USA}
\affiliation{Center for Cosmology and AstroParticle Physics (CCAPP), The Ohio State University, Columbus, OH 43210, USA}
\affiliation{Research School of Astronomy \& Astrophysics, Australian National University, Cotter Rd., Weston, ACT 2611, Australia}
\affiliation{School of Computing, Australian National University, Acton, ACT 2601, Australia}

\begin{abstract}

We report on a data-driven spectral model that we have developed for the identification of double-lined spectroscopic binary stars (SB2s) in the LAMOST low-resolution survey (R$\sim$1800). Employing simultaneous fitting with both single-star and binary-star models, we detected over 4800 SB2 candidates, where both components are detectably contributing to the spectrum, from an initial pool of 2.6 million objects. Tests show that our model favors FGK-type main-sequence binaries with high mass ratio ($q\geq$ 0.7) and large radial velocity separation ($\Delta \rm RV \geq$ 100~km$\,$s$^{-1}$). Almost all of these candidates are positioned above the main sequence in the color-magnitude diagram, indicating their binary nature. Additionally, we utilized various observational data, including spectroscopy, photometry, parallax, and extinction, to determine multiple physical parameters such as the effective temperature, age, metallicity, radial velocity, mass, mass ratio, stellar radius, along with their associated uncertainties for these SB2 candidates. For the 44 candidates with seven or more observational epochs, we provide complete orbital solutions. We make available catalogs containing various stellar parameters for identified SB2 systems.
\end{abstract}


\section{Introduction}\label{sec:intro}

Approximately half of the solar-type stars exist within binary or more complex multiple-star systems \citep{2010ApJS..190....1R, 2013ARA&A..51..269D, 2017ApJS..230...15M, 2021A&A...650A.201R}. Investigating binary systems holds significance not only for comprehending their formation, evolution, and interactions but also for advancing in other scientific domains, primarily
 including the compact object binaries (e.g., \citealt{2019Sci...366..637T,2019Natur.575..618L,2020ApJ...900...42L,2022NatAs...6.1203Y,2023MNRAS.518.1057E,2023MNRAS.521.4323E,2024A&A...686L...2G}) and Type Ia supernova progenitors (e.g., \citealt{1984ApJ...277..355W,2006MNRAS.368.1095H,2011NatCo...2..350H,2012NewAR..56..122W}).

Based on different detection methods, binary stars can roughly be divided into visual binaries and spectroscopic binaries (SBs). Spectroscopic binaries (SBs) are further differentiated into single-lined binaries (SB1s) and double-lined binaries (SB2s), depending on whether one or both stellar components are observable in their spectra. SB1s can be identified through the detection of radial velocity (RV) variability \citep{2010AJ....140..184M,2020ApJS..249...31Y,2022NatAs...6.1203Y,2022ApJ...933..193Z,2022ApJ...936...33Z,2023SCPMA..6629512Z}, whereas SB2s are mainly detected using cross-correlation functions (CCF; \citealt{2017A&A...608A..95M, 2021AJ....162..184K, 2021ApJS..256...31L, 2023arXiv231204721V}), machine learning methods \citep{2018MNRAS.476..528E, 2020A&A...638A.145T, 2022ApJS..258...26Z} and projected rotational velocities \citep{2022MNRAS.517..356K}. Significant efforts have been made to identify and study SBs through high or medium-resolution stellar spectroscopic surveys, such as the Apache Point Observatory Galactic Evolution Experiment (APOGEE; \citealt{2017AJ....154...94M,2017PASP..129h4201F,2018MNRAS.476..528E,2020ApJ...895....2P,2021AJ....162..184K}), the Large Sky Area
Muti-Object fiber Spectroscopic Telescope Medium-Resolution Survey (LAMOST MRS; \citealt{2012RAA....12..723Z, 2019RAA....19...64Q, 2021ApJS..256...31L, 2022ApJS..258...26Z, 2022MNRAS.517..356K}), the Galactic Archaeology with HERMES (GALAH; \citealt{2015MNRAS.449.2604D, 2020A&A...638A.145T}), and the Gaia-ESO Survey (GES; \citealt{2022A&A...666A.120G, 2022A&A...666A.121R, 2017A&A...608A..95M, 2023arXiv231204721V}). Moreover, $Gaia$ Data Release 3 ($Gaia$ DR3; \citealt{2023A&A...674A...1G}) has provided two-body orbital solutions for astrometric, spectroscopic, and eclipsing binaries, which significantly expands the known binary orbits, offering a valuable resource of dynamical masses and making a significant contribution to the analysis of stellar multiplicity.

However, few studies have focused on searching for SB2s in low-resolution spectroscopic (e.g., $R\sim1800$) surveys which usually contain data for a much larger number of sources. The LAMOST Low-Resolution Survey (LRS; \citealt{2015RAA....15.1095L}), beginning in November 2011, is one of the largest stellar spectroscopic surveys to date, covering the wavelength range of 3700-9000~\AA\ with a spectral resolution of $R\sim1800$.
The LAMOST LRS data release 9 (DR9)\footnote{http://www.lamost.org/dr9/v1.0/} released 11,226,252 spectra for 10,907,516 stars. The scale of the LAMOST LRS DR9 makes it an ideal database for mining SB2 systems.

In this work, we employ a spectral fitting method to identify SB candidates in LAMOST LRS DR9 and to estimate their stellar parameters. We introduce the data processing strategy, including model construction and data preprocessing, in Section~\ref{sec:methods}. In Section~\ref{sec:Detection_Efficiency}, we set the selection criteria for binary candidates and certify them. Section~\ref{sec:Result&Validation} describes how we validate these binary candidates using photometric data and determine their orbital parameters through dynamical methods. We compare our findings with other catalogs and summarize our results in Sections~\ref{subsec:ComparisonOtherCatalogs} and \ref{Summary}, respectively.

\begin{figure}
\centering
\includegraphics[trim=0 0 0 0, clip, width = 8.5cm]{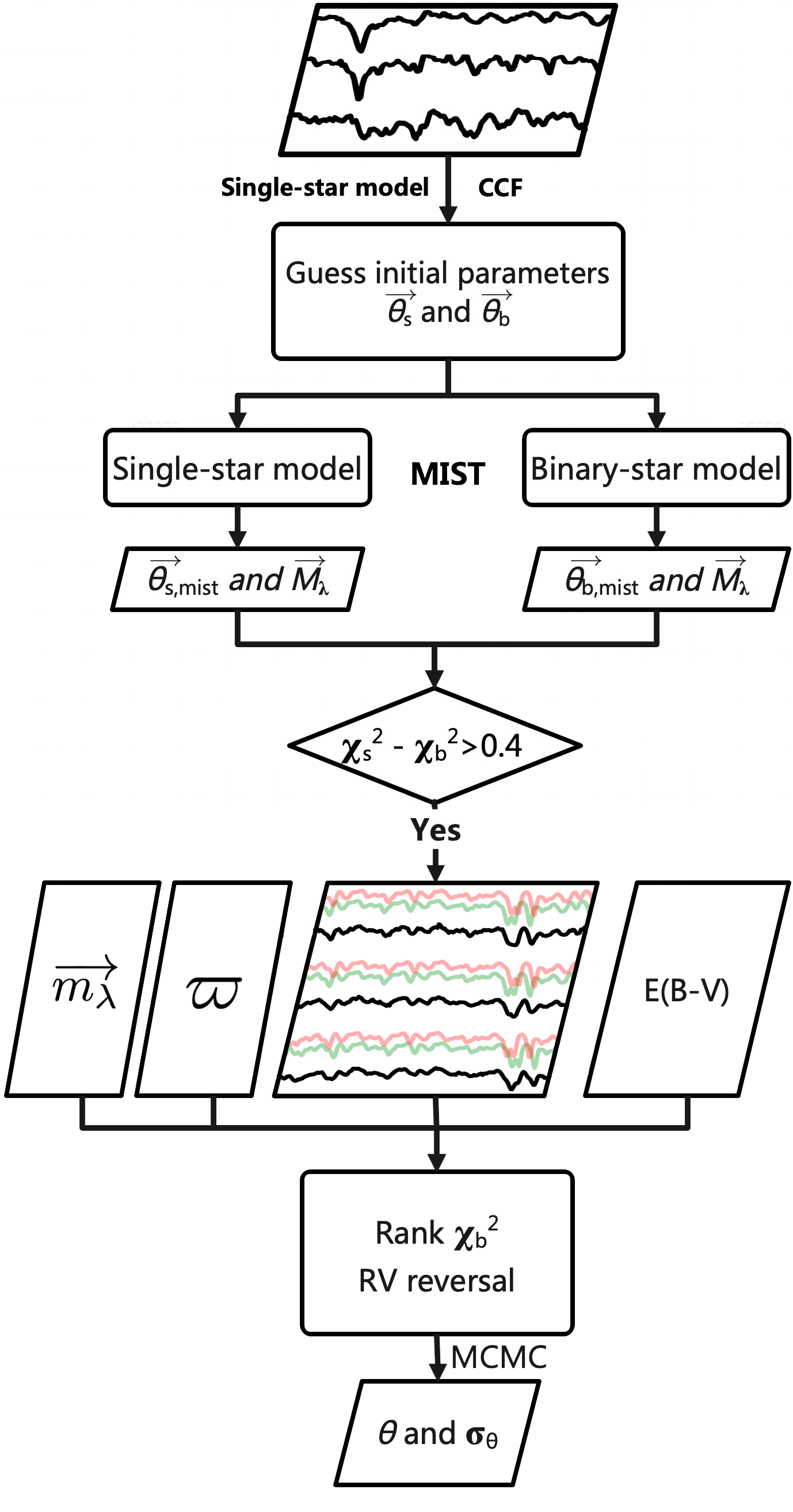}
\caption{The flow chart of this work. The main processes include initial parameter estimation (Section~\ref{sec:preprocessing}); single-star and binary-star models fitting (Section~\ref{sec:Spectral_model_fitting}); determination of binary star candidates ($\chi^{2}_{\rm s} - \chi^{2}_{\rm b} > 0.4$; Section~\ref{subssec:Find_the_criteria}); RV reversal (Section~\ref{subsec:reverse_RV}) considering spectroscopic, photometric ($\overrightarrow{m}_{\rm \lambda}$), extinction (E(B-V)), and astrometric ($\varpi$) information, followed by MCMC fitting (Section~\ref{subsec:MCMC}); and finally, obtaining the parameters and their uncertainties for binary star candidates (Table~\ref{table:4848 binary candidates}).}
\label{fig:flow_chart}
\end{figure}

\section{Methods} \label{sec:methods}

We employ the following steps for spectral binary candidates searching: (i) training a single-star model based on the observational spectra, and conducting model validation (Sections~\ref{subsubsec:LAMSOTdata}--\ref{sec:Single-star-model}); (ii) generating a binary-star model by combining two single-star models coupled with RVs (Section~\ref{sec:Binary-star-model}); (iii) single-star and binary-star models fitting (Section~\ref{sec:Spectral_model_fitting}); (iv) experimental criteria for SB2 candidates identification (Section~\ref{subssec:Find_the_criteria}). Upon the identifications, we utilize the method of RV reversal to allocate RVs as accurately as possible to the correct component stars in each observation (Section~\ref{subsec:reverse_RV}). Meanwhile, we integrated spectroscopic, photometric, extinction, and astrometric information, employing the Markov Chain Monte Carlo (MCMC) technique, to ascertain stellar parameters while simultaneously generating uncertainty estimates for each parameter (as detailed in Section~\ref{subsec:MCMC}). For optimizing the nonlinear least-squares fitting between the observed and model spectra, we employed the $\tt least~squares$ algorithm in $\tt SciPy$\footnote{https://docs.scipy.org/doc/scipy/index.html}. A binary candidate can be identified when a binary-star model can significantly better fit ($\Delta \chi^{2} > 0.4$; see Sections~\ref{sec:Spectral_model_fitting}) the spectra of this object than a single-star model. Figure~\ref{fig:flow_chart} shows the flow chart of this work, which will be elucidated in detail in the following sections. In addition, the methodology of this work has been applied to \citet{2024ApJ...964..101Z} and \citet{2024ApJ...965..167L}.

\begin{figure*}
\centering
\includegraphics[width = 18.5cm]{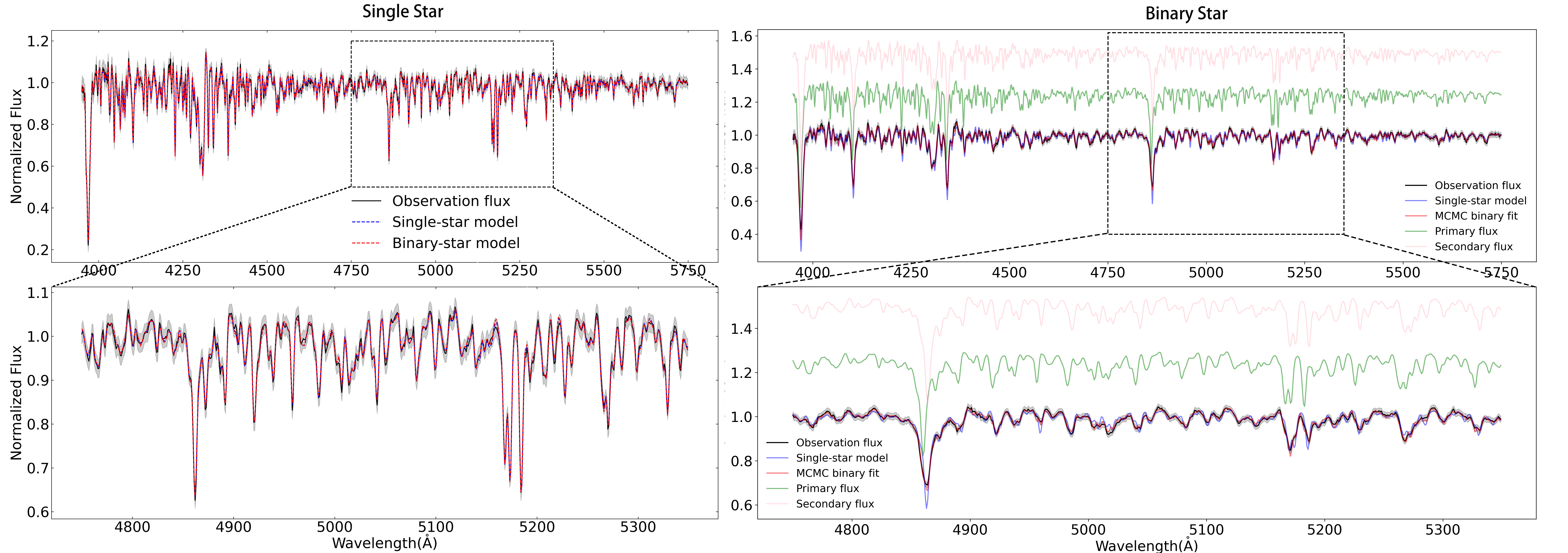}
\caption{Comparing the spectral fittings with single-star (blue) and binary-star (red) models. The upper panels show the complete normalized spectra with the best-fit two models, while the lower panels illustrate the zoomed-in view of the spectra for a certain wavelength range. Left: Both models show an almost identical fit to the spectrum. Right: The binary-star model fitting is significantly better than that of the single-star model. The spectra of the primary (pink) and secondary (green) stars are also shown.}
\label{fig:Compare_single&binary_model}
\end{figure*}

In Figure~\ref{fig:Compare_single&binary_model}, we demonstrate the process of determining whether a spectrum is from a single star or a binary system by comparing the fitted spectra with both the single-star and binary-star models and the observed spectrum simultaneously. The left panel represents that both the single-star (Section~\ref{sec:Single-star-model}) and binary-star (Section~\ref{sec:Binary-star-model}) models are consistent with the observed spectrum and produce similar fitting results. The right one shows a binary candidate with $\Delta \chi^{2} > 0.4$, showcasing a remarkably improved fit by the binary-star model compared to the single-star one. Especially, there are some distinctive absorption line features resulting from the orbital motion of the binary, such as the $\rm H \beta$ at 4862.68 \AA\ and the Mg I triplet at 5167, 5172, and 5183 \AA, which cannot be reproduced by any single-star model.

\subsection{Data pre-processing} \label{sec:preprocessing}

The CCF method \citep{1979AJ.....84.1511T} is a widely used approach for calculating RV by cross-correlating the observed spectra with the best-matched template. It provides us with the RV, as well as the spectral parameters (namely, effective temperature $T_{\rm eff}$, surface gravity $\log g$, metallicity [M/H], and [$\alpha$/M]) of the best-matched template spectra. These parameters enable us to pre-filter the raw data from LAMOST LRS DR9 with g-band signal-to-noise ratio (SNR) $\geq\,30$ (4,289,736 spectra). 

\begin{figure}
\centering
\includegraphics[width = 12cm]{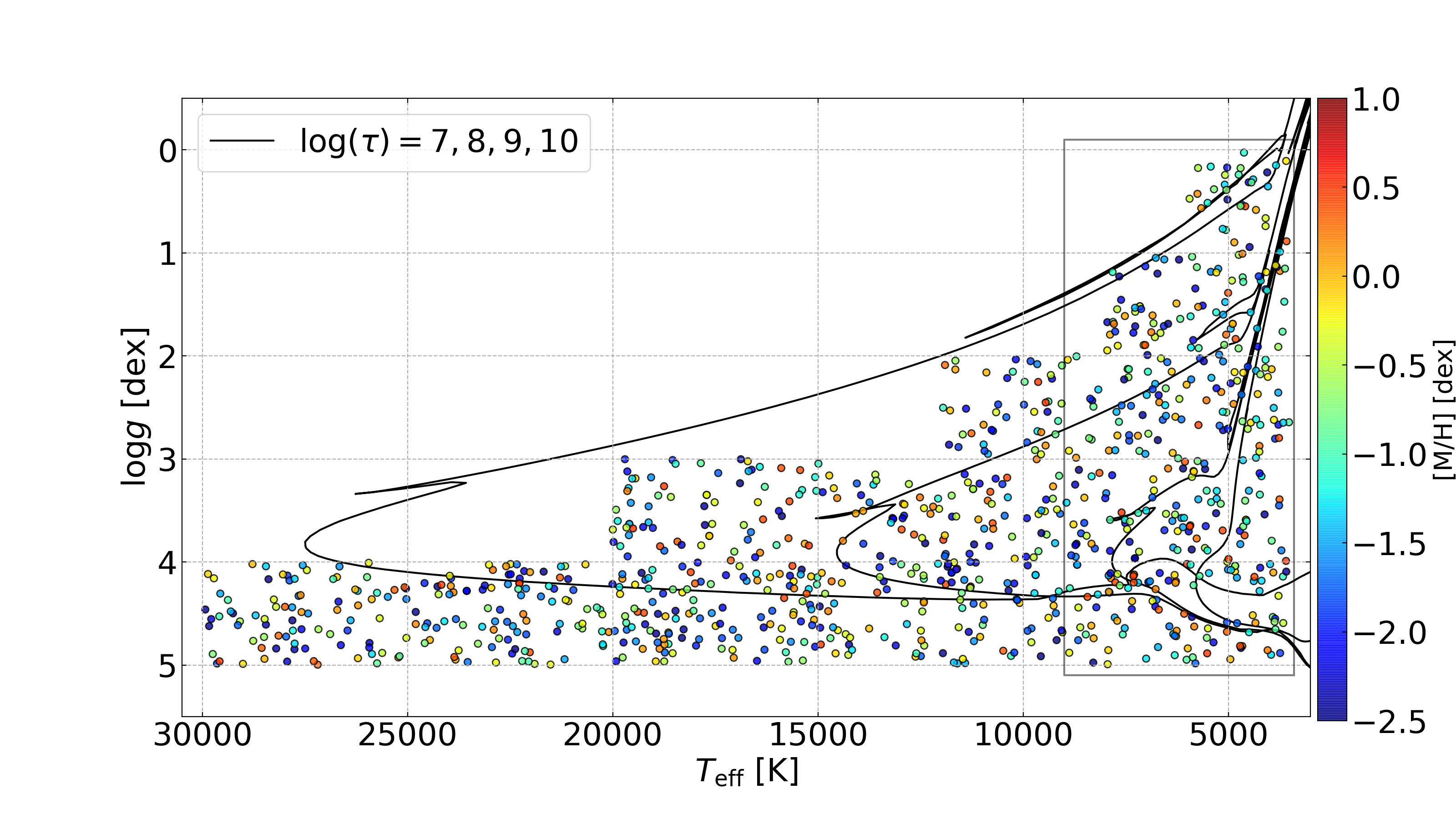}
\caption{The distribution of $T_{\rm eff}$, $\log g$ and [M/H] of CCF template spectra. Superposed is a series of isochrones with solar metallicity but different ages $\tau$ in years. The gray box indicates the parameter space with $ T_{\text{eff}} < $9000K left in the data pre-processing using the CCF method, which includes 405 template spectra.}
\label{fig:CCF_template}
\end{figure}

Our template comprises 1000 synthetic spectra, uniformly distributed across the parameter range:

\begin{enumerate}
    \item 3500~K  $< T_{\rm eff} <$ 30000~K,
    \item 0.0 $<\log g < 5.0$ dex,
    \item $-2.5 <$ [M/H] $< 1.0$ dex,
    \item $-1.0 <$ [$\alpha$/M] $< 1.0$ dex.
\end{enumerate}
The distribution of these parameters is depicted in Figure~\ref{fig:CCF_template}, with the the parameter space retained by the CCF shown within the gray box.

After removing high-temperature sources with $T_{\rm eff}>9000$~K using CCF, we applied the single-star model (Section~\ref{sec:Single-star-model}) to all samples and mainly excluded giant stars with $\log g<3.5$. Following these two screenings, our dataset comprises 3,584,214 spectra, for 2,027,146 objects each with one single observation, and 592,622 objects with multiple observations (1,557,068 spectra in total).

\begin{figure*}
\centering
\includegraphics[width = 13cm]{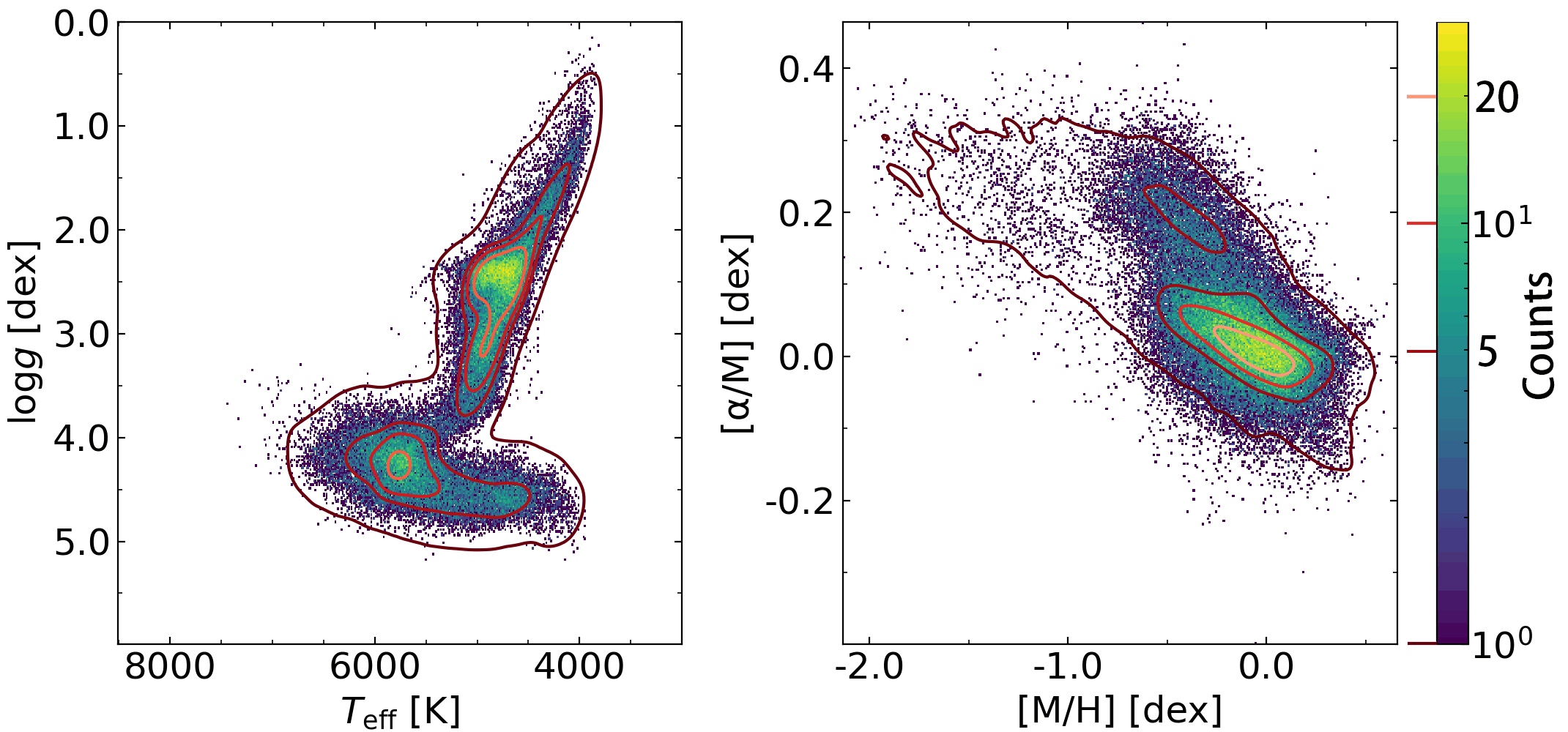}
\caption{The distribution of $T_{\rm eff}$ versus $\log g$ (left panel), and [M/H] versus [$\alpha$/M] (right panel). The heatmaps represent the distribution of APOGEE parameters, while the contour plots depict the distribution of our model's predicted parameter values.}
\label{fig:Training}
\end{figure*}

\subsection{Training Data Sample}\label{subsubsec:LAMSOTdata}
The key of APOGEE Stellar Parameter and Chemical Abundance Pipeline (ASPCAP; \citealt{2016AJ....151..144G}) is the utilization of the FERRE program \citep{2006ApJ...636..804A} to align the observed spectrum with an interpolated grid of synthetic spectra, thereby deriving the parameters of the best-fitting template. By employing a cross-match radius of 3\arcsec, we establish optical-to-optical correspondences between the spectra of LAMOST LRS DR9 and spectral parameters of APOGEE catalogs. The initial training set, comprising 74,768 objects, is assembled by applying the following criteria:
\begin{enumerate}
    \item 3500~K $< T_{\rm eff} <$ 8000~K,
    \item $0<\log g < 5.0$\,dex, 
    \item $-2.0<$ [M/H] $<1.0$\,dex,
    \item $|V_{\rm APOGEE} - V_{\rm LAMOST}| < 50$ km~s$^{-1}$,
    \item ``ASPCAPFLAG" = 0,
    \item the SNR of APOGEE spectra and that of LAMOST LRS DR9 at the $g$-band are all greater than 30, 
\end{enumerate}
where $T_{\rm eff}$ is the effective temperature, $\log g$ is the surface gravity, [M/H] is the overall metallicity, $V_{\rm APOGEE}$ and $V_{\rm LAMOST}$ are the observed RVs in APOGEE and LAMOST, respectively, and ``ASPCAPFLAG" is the informational bitmask\footnote{https://www.sdss4.org/dr17/irspec/aspcap/} in APOGEE DR16. The first three criteria define the parameter ranges for the stellar spectral fitting. The fourth criterion excludes objects potentially associated with multi-body systems, where RV shifts are detected in different observations. The last two criteria ensure that the selected sample spectra exhibit sufficiently high SNRs and reliable parameter values in both the APOGEE and LAMOST catalogs. Within this training set, we establish a direct coupling of labels from APOGEE DR16, including $T_{\rm eff}$, $\log g$, [M/H], and overall $\alpha$-element abundance [$\alpha$/H], with the corresponding spectra flux data from LAMOST LRS DR9.

\subsection{Single-star model}\label{sec:Single-star-model}

The Stellar LAbel Machine\footnote{https://github.com/hypergravity/laspec} (SLAM; \citealt{2020ApJS..246....9Z, 2020RAA....20...51Z}) is a neural network-based forward stellar model that employs the robust nonlinear regression technique of the support vector regression (SVR; \citealt{1995Support}) algorithm. It serves as the training framework for all models in this study.

Firstly, we perform the normalization of the LAMOST LRS DR9 spectra using a low-order polynomial with the $\tt normalization$ function in the $\tt LASPEC$ package \citep{2020ApJS..246....9Z}. We input the normalized spectra in the wavelength range of 3950--5750 \AA\ and the stellar parameter vectors $\overrightarrow{\theta}$ as ingredients to the $\tt SLAM$ for training the single-star model $\phi$. In essence, we train the LAMOST LRS DR9 spectra as a data-driven generative model $\phi$ that can predict the rest-frame normalized flux density at a given wavelength. This prediction is achieved as a function of a set of `labels', denoted by $\overrightarrow{\theta}$, which determine the characteristics of the spectrum. This model $\phi$ equips us with the ability to convert normalized spectra to stellar parameter vectors $\overrightarrow{\theta}$, and vice versa. The vector $\overrightarrow{\theta}$ of a single star is 
\begin{equation}\label{equ:phi_train}
    \overrightarrow{\theta} = \lbrace T_{\rm eff}, \log g, [\rm M/\rm H], [\alpha/\rm M] \rbrace,
\end{equation}
where the [\rm M/\rm H] and [$ \rm \alpha$/\rm M] are overall metallicity and $\alpha$-element abundance in APOGEE DR16, respectively. Then, we can employ the single-star model to predict a normalized spectrum with no radical velocity for a certain $\overrightarrow{\theta}$ as,
\begin{equation}\label{equ:flux_single}
    \overrightarrow{f_{\rm s}} = \phi(\overrightarrow{\theta}).
\end{equation}

\begin{figure*}
\centering
\includegraphics[width = 19cm]{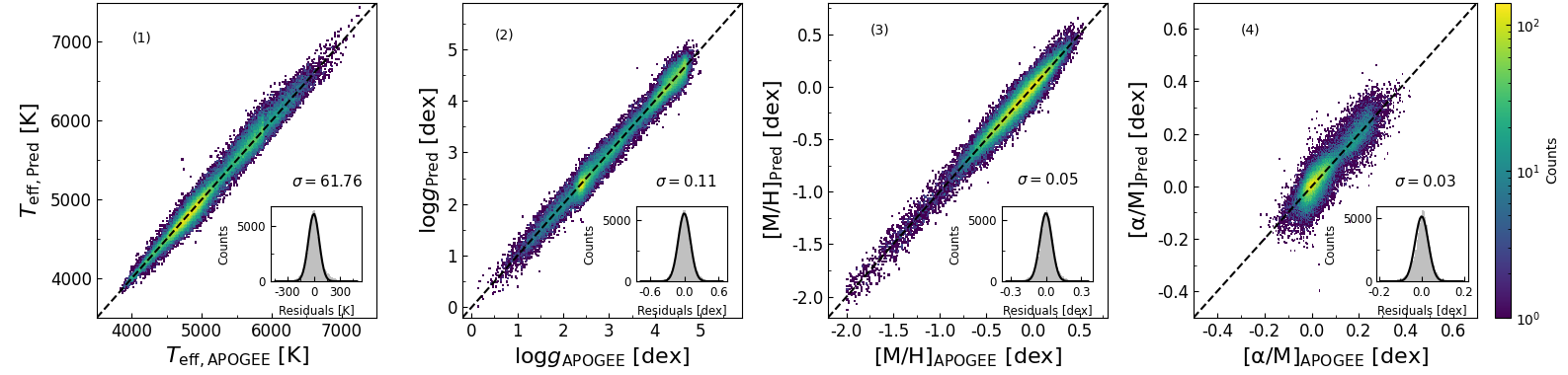}
\caption{The comparison distributions of $T_{\rm eff}$, $\log g$, [M/H] and [$\alpha/M$] values from our predictions and APOGEE labels. A Gaussian fit to the residuals is shown in the lower-left inset of each plot.}
\label{fig:label_compare}
\end{figure*}

The final single-star model is the result of an iterative refinement process designed to eliminate portions of the spectra that cannot be accurately predicted. This refinement involves applying the training model to its original dataset to generate a prediction vector, denoted as $\overrightarrow{\theta}_{\rm pred}$. We then calculate the discrepancies between this prediction vector and the input vector from APOGEE. To identify and remove problematic samples, we employ a Gaussian function to model the distribution of these discrepancies, eliminating any samples whose parameter distribution falls outside a $4\sigma$ threshold. Through three rounds of iterative screening and retraining, the original set of 74,768 objects was reduced to 64,013. The comparisons between the predicted labels (contours) and APOGEE labels (heatmap) for both $T_{\rm eff}$ versus $\log g$ (left panel) and [M/H] versus [$\alpha$/M] (right panel) are illustrated in Figure~\ref{fig:Training}. These comparisons highlight that the distribution and density of predicted labels from our model closely match those from APOGEE. Furthermore, Figure~\ref{fig:label_compare} presents a detailed comparison of each predicted parameter with the corresponding APOGEE parameters. The uncertainties associated with these four parameters are as follows: $\sigma(T_{\rm eff}) = 61.76~\rm K$, $\sigma(\log g) = 0.11~\rm dex$, $\sigma([\rm M/H]) = 0.05~\rm dex$ and $\sigma([\rm \alpha/M]) = 0.03~\rm dex$. These values suggest that the parameters predicted by our single-star model are closely consistent with the stellar parameters of the pre-labeled APOGEE objects.

In our actual spectral fitting, the vector of free parameters of a single star with RV is 
\begin{equation}\label{equ:theta_single}
    \overrightarrow{\theta_{\rm s}} = \lbrace m, \rm log \tau, [\rm M/\rm H], [\alpha/\rm M], \textit{v}_{\rm i} \rbrace,
\end{equation}
where $m$ is the stellar mass, $\tau$ is its age in years, and $v_{\rm i}$ is the RV in the $i$-th observational epoch.
Then we use the MIST stellar evolutionary model trained from stellar evolutionary tracks \citep{2016ApJS..222....8D, 2016ApJ...823..102C} to convert $\overrightarrow{\theta_{\rm s}}$ into the atmospheric parameters of the single star, i.e., 
\begin{equation}\label{equ:phi_single}
    \overrightarrow{\theta}_{\rm s, mist} = \lbrace T_{\rm eff}, \rm \log \textit{g}, [\rm M/\rm H], [\alpha/\rm M], \textit{v}_{\rm i} \rbrace.
\end{equation}
Meanwhile, the radii $R$ can be derived by the MIST model, which is vital in the calculation of the luminosity ratio of two components in binary systems. The $\phi (\overrightarrow{\theta}_{\rm s, mist})$ is the predicted normalized flux of a single star considering RV.

\subsection{Binary-star model}\label{sec:Binary-star-model}

Assuming the same age and stellar composition, the two components of a binary star share the identical values of age, $[\rm M/\rm H]$ and [$\alpha$/\rm M], then the vector containing free parameters for a binary object is 
\begin{equation}\label{equ:theta_binary}
    \overrightarrow{\theta_{\rm b}} = \lbrace m_{1}, q, \rm log \tau, [\rm M/\rm H], [\alpha/\rm M], \textit{v}_{1,i}, \gamma \rbrace,
\end{equation}
where $m_{1}$ is the mass of the primary component; $q = m_{2} / m_{1}$ ($m_{2} \leq m_{1}$) is mass ratio while $m_{2}$ is the mass of the secondary component; $v_{1,\rm i}$ is the observed RV of the primary star for its $i$-th observation; $\gamma$ is the center-of-mass heliocentric velocity of binary. For a binary system, then the observed RV of the secondary star is $v_{2,\rm i} = (\gamma \times (q+1)-v_{1,\rm i})/q$. 

By adopting the MIST model, we further convert $\overrightarrow{\theta_{b}}$ to the stellar parameter for primary and secondary respectively, i.e.,
\begin{equation}\label{equ:phi_primary}
    \overrightarrow{\theta}_{\rm b,mist,1} = \lbrace T_{\rm eff, 1}, \log g_{1}, [\rm M/\rm H], [\alpha/\rm M], \textit{v}_{1,\rm i} \rbrace.
\end{equation}
and 
\begin{equation}\label{equ:phi_secondry}
    \overrightarrow{\theta}_{\rm b,mist,2} = \lbrace T_{\rm eff, 2}, \log g_{2}, [\rm M/\rm H], [\alpha/\rm M], \textit{v}_{2,\rm i} \rbrace.
\end{equation}
where $T_{\rm eff, 1}$ and $\log g_{1}$ represent the effective temperature and surface gravity of the primary star, respectively, while $T_{\rm eff, 2}$ and $\log g_{2}$ denote the corresponding values for the secondary star.

Different from the single-star model, which only requires consideration of the normalized spectrum, the binary star model needs to incorporate the non-normalized spectra of both the primary and secondary stars. The nature of a non-normalized spectrum is a normalized single-star spectrum multiplied by the corresponding model-predicted pseudo-continuum.

To create the pseudo-continuum model, we employed the code $\tt Regli$ \citep{https://doi.org/10.5281/zenodo.3461514} to generate a series of theoretical spectra in grids defined as in Section~\ref{sec:preprocessing}, which are non-normalized spectra containing the flux value of each spectrum at each wavelength. We extracted 30,000 pseudo-continuum spectra and their corresponding labels as the same as $\overrightarrow{\theta}$. Then, adopting the same training process as single stars in Section~\ref{sec:Single-star-model}, we obtained the pseudo-continuum spectral model $\psi$, which is utilized to compute the corresponding pseudo-continuum spectra based on $\overrightarrow{\theta}_{\rm b, mist,1}$ and $\overrightarrow{\theta}_{\rm b,mist,2}$, facilitating the development of the binary-star model.

The non-normalized binary-star spectrum with two components is 
\begin{equation}\label{equ:STW_LX_P}
    \overrightarrow{F_{b}} = \overrightarrow{f_{1}} \times R_{1}^{2} \times \psi(\overrightarrow{\theta}_{\rm b,mist,1}) + \overrightarrow{f_{2}} \times R_{2}^{2} \times \psi(\overrightarrow{\theta}_{\rm b,mist,2}). 
\end{equation}
where $\overrightarrow{f_{1}}=\phi(\overrightarrow{\theta}_{\rm b,mist,1})$ and $\overrightarrow{f_{2}}=\phi(\overrightarrow{\theta}_{\rm b,mist,2})$ represent the normalized spectra of the primary and secondary stars, respectively; $R_{1}$ and $R_{2}$ are the radii derived from the MIST and $\psi(\overrightarrow{\theta}_{\rm b})$ is the pseudo-continuum. The $\overrightarrow{F_{\rm b}}$ is then normalized to get $\overrightarrow{f_{\rm b}}$ as the binary-star model.

\subsection{Spectral model fitting}\label{sec:Spectral_model_fitting}

We employed the \texttt{SCIPY least$\_$squares} with the method `Trust Region Reflective' algorithm \citep{Branch1999ASI} to carry out the model fitting. The $\chi^{2}$ was utilized to characterize the difference between the fitted results and the observational spectra for both single-epoch and multiple-epochs sources, as follows,
\begin{equation}\label{equ:chi2}
    \chi^{2} = \sqrt{(\frac{\overrightarrow{f}_{\rm fit} - \overrightarrow{f}_{\rm obs}}{N \times \overrightarrow{\sigma}_{\rm f}})^{2}}, 
\end{equation}
where $\overrightarrow{f}_{\rm fit}$ and $\overrightarrow{f}_{\rm obs}$ are fitting and observed 
normalized spectra respectively, $N$ is the number of spectral wavelength, and $\overrightarrow{\sigma_{f}}$ is the error of observed spectra. Hence, the primary objective of the model fitting process is the minimization of the $\chi^{2}$ values, denoted as $\chi^{2}_{s}$ for the single-star model, and $\chi^{2}_{b}$ for the binary-star model. In the single-star model fitting, the initial RV values were set to correspond with the RVs from CCF. When fitting the binary-star model, we employed the same methodology as outlined in Section~2.3 of \citet{2018MNRAS.476..528E} to initialize the $q$ values. Within a range spanning plus or minus 50 km~s$^{-1}$ around the CCF's RV values, we uniformly selected 10 initial values for $v_{\rm 1,i}$, while $\gamma$ was sampled with 10 initial values falling within the -150 km~s$^{-1}$ to 150 km$\,$s$^{-1}$ range. Ultimately, we selected the fitting result associated with the lowest $\chi^{2}$ value. Through the ‘semi-empirical’ synthetic spectral experiments detailed in Section~\ref{subssec:Find_the_criteria}, we ultimately utilized the criterion of $\Delta\chi^{2} = \chi^{2}_{\rm s} - \chi^{2}_{\rm b} > 0.4$ to quantitatively identify binary star candidates, where $\Delta\chi^{2}$ is the differences between the single-star and binary-star model fittings.

\subsection{Find the criterion for identifying binary} 
\label{subssec:Find_the_criteria}

\subsubsection{Mock binary star spectral fitting} \label{subsec:mock_binary}

To access the accuracy and system properties of our method, and to measure the expected $\Delta \rm \chi ^{2}$ values for true binary at specific mass ratios, we constructed a library of 10,000 `semi-empirical' synthetic binary-star spectra with the spectral resolution of LAMOST LRS DR9 by employing the same method of synthesizing the binary star spectra in Section~\ref{sec:Binary-star-model}. The stellar parameters of these spectra are based on the training binary library from \citet{2022ApJS..258...26Z}, except that we set the $\log g>3.5$ and temperature between 3500~K and 8000~K. The detailed parameters are listed in Table~\ref{table:Test_mock_binary_parameters}. We do not attempt to fit the spectra of giants. Because in most giant-dwarf binary systems, the smaller dwarf star contributes very little to the overall brightness, making it difficult to distinguish. In giant-giant binary systems, both components typically have similar masses and, therefore, exhibit quite similar spectra, especially if they are of the same age \citep{2018MNRAS.476..528E}. The SNRs of the synthetic spectra, ranging greater than 30, are drawn from an identical SNRs distribution of LAMOST LRS DR9.

 \begin{table}
   \caption{The summary of the parameters of the mock binary star.}
     \label{table:Test_mock_binary_parameters}
   \begin{center}
   \begin{tabular}{lcc}\hline \hline

Quantity & Distribution & Unit \\
\hline
$m_{1}$ & $ p(m_{1})=\left\{\begin{aligned} 1 && 0.1 < m_{1} < 1 \\ m_{1}^{-2.35} && 1 < m_{1} < 2 \end{aligned} \right.$ &  M$_{\odot}$\\
$q$ & (0.001, 1) & \\
$\log \rm g$ & (3.5, 5.0) & dex \\
$[$M/H$]$ & (-2.0, 0.5) & dex \\
$[\alpha$/M$]$ & (-0.5, 0.7) & dex \\
$v_{1}$ & (-300, 300) & km s$^{-1}$ \\
$\Delta v$ & (-500, 500) & km s$^{-1}$ \\
\hline\noalign{\smallskip}
  \end{tabular}
  \end{center}
\end{table}

\begin{figure*}
\centering
\includegraphics[trim=0 0 0 0, clip, width = 20cm]{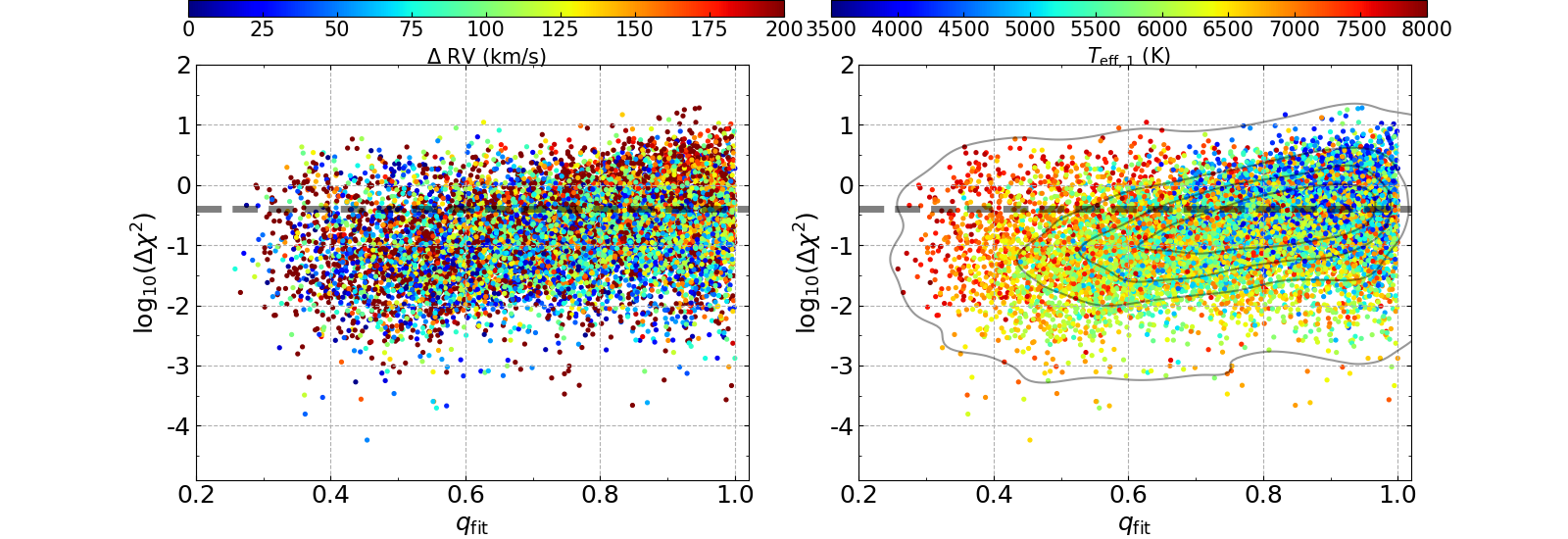}
\caption{The $q$ versus $\log (\Delta \chi^{2})$ from the results of single-star and binary-star model fitting for the semi-empirical binary spectra. The color bar on the left is the RV offset $\Delta \rm RV$ between the two components and the one on the right is the temperature of the primary star. The dashed lines locate at $\log(\Delta \chi^{2}) = -0.4$. The black contour lines in the right panel represent the relative density of the data points. The closer to the inner rings, the higher the density of the data points.}
\label{fig:D_chi2}
\end{figure*}

The $q$ versus $\Delta \chi^{2}$ distributions are shown in Figure~\ref{fig:D_chi2}. The color bars plotted in the left and right panels are 
the RV offsets between two component stars ($\Delta \rm RV$) and the temperature of the primary star ($T_{\rm eff,1}$), respectively. In this experiment, 9652 mock binaries can be simply filtered out by $\Delta \chi^{2} > 0$.
As shown in the left panel, $\Delta \chi^{2}$ is a good function of $q$ for objects with $q\geq\,$0.7, i.e., the greater the $\Delta$RV, the greater the $\Delta \chi^{2}$. For a binary star with $0.7\leq q \leq 1.0$, the luminosity contribution of the secondary star is close to that of the primary. For a binary system with $q\geq0.7$, the larger RV offset makes it easier to distinguish the two components. In addition, as expected, binaries with a hotter primary have a wider range of $q$.

\begin{figure*}
\centering
\includegraphics[trim=10 0 0 0, width = 18cm]{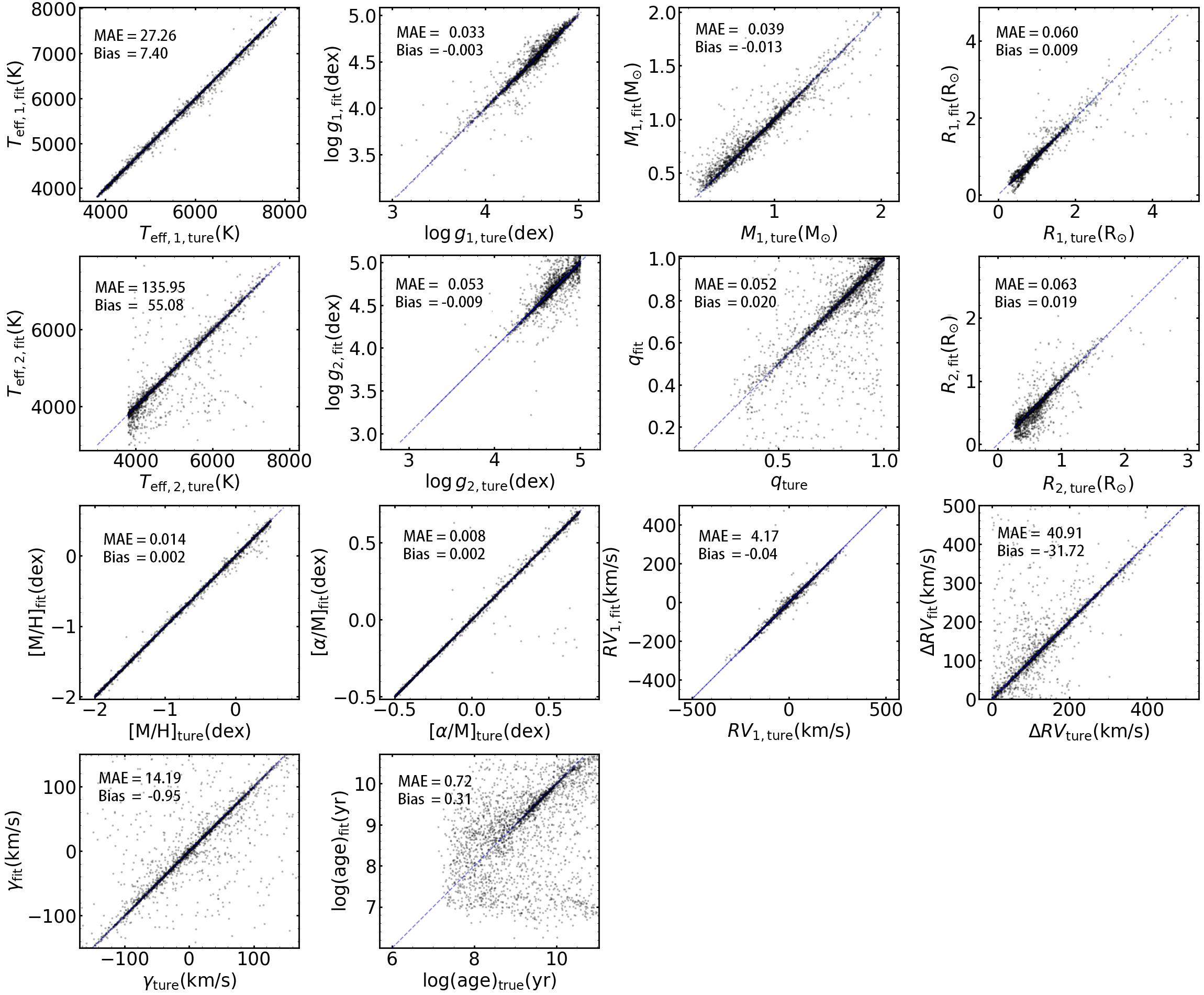}
\caption{The comparison distribution of different stellar and dynamical parameters from binary-star model fitting and true labels. For each label $\mathcal{L}_{i}$, the MAE is the mean absolute error, defined as mean($|\mathcal{L}_{i, true} - \mathcal{L}_{i, fit}|$), and the bias, defined as mean($\mathcal{L}_{i, true} - \mathcal{L}_{i, fit}$).}
\label{fig:one-to-one_plot}
\end{figure*}

\begin{figure}
\centering
\includegraphics[width = 8cm]{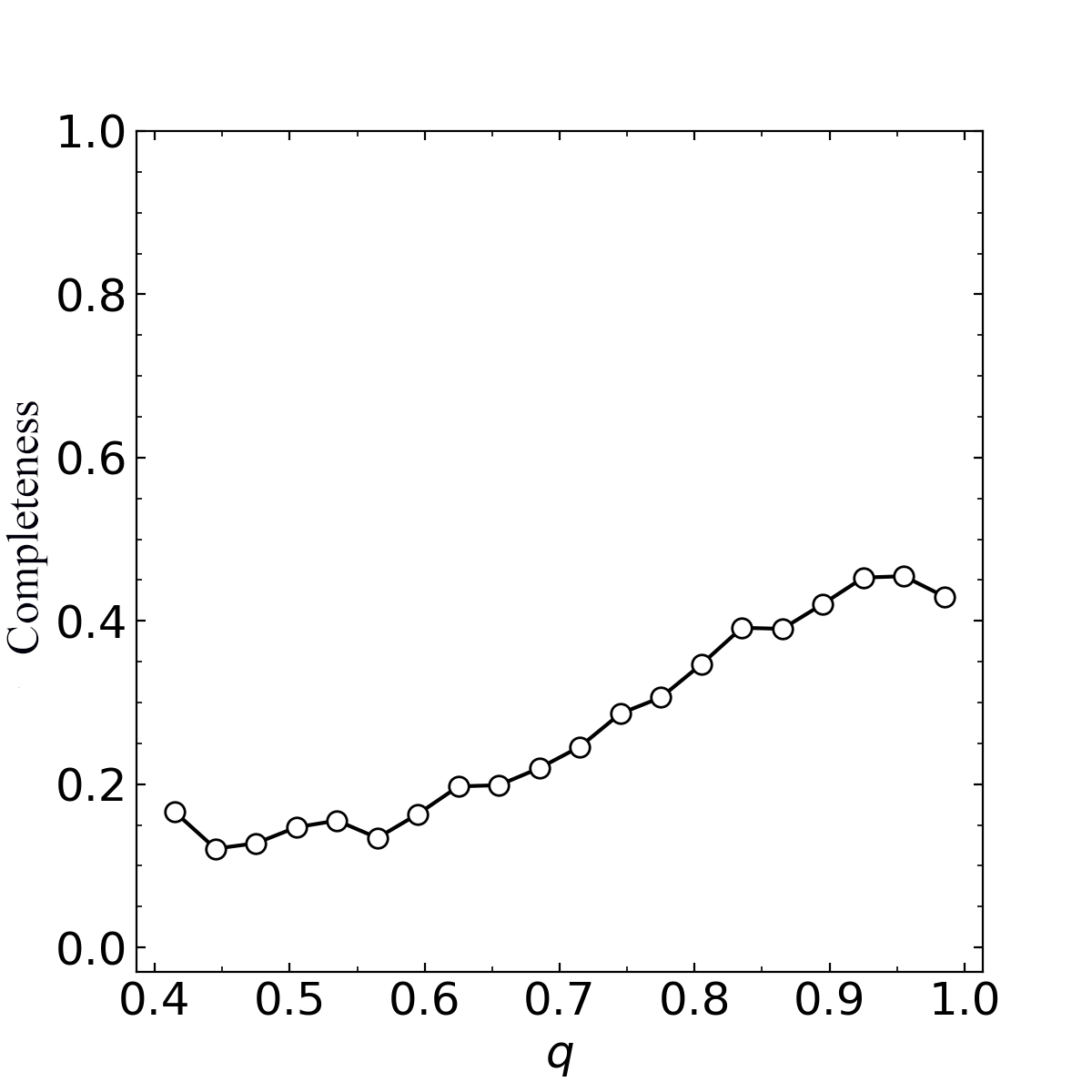}
\caption{The completeness of the binary identifications in this work. The x-axis represents the mass ratio, while the y-axis indicates the proportion of mock binaries that meet the threshold $\Delta \chi^{2} > 0.4$ at each mass ratio.}
\label{fig:completeness}
\end{figure}

Based on the experiments in Section~\ref{subsubsubsec:mock_bianry_9000}, we find that $\Delta \chi^{2} > 0.4$ (i.e., $\log\Delta \chi^{2} = -0.4$) is a proper criterion for identifying binary candidates in this work. 
After adopting this criterion as the binary detection threshold, 2916 out of the 10,000 can be identified as binary stars by our method. Then, we compare the predicted values of our model with the true values of the semi-empirical spectra, as shown in Figure~\ref{fig:one-to-one_plot}. We calculated the values of mean absolute error (MAE) and bias for each parameter, revealing that the distributions of different parameters closely align with the one-to-one lines with small scatter. This affirms the method's reasonable precision in predicting binary star parameters. The scatter of parameters for the secondary star (e.g., temperature, surface gravity) is slightly higher than that of the primary star. This is attributed to the compounded errors in the calculation of the secondary star's parameters, arising from mass ratio and age fitting. The completeness of binary stars along $q$ of this work can be derived and shown in Figure~\ref{fig:completeness}. As the mass ratio increases, the completeness of the detected binary stars gradually increases. Owning to the resolution of LRS, the highest completeness is $\sim$0.45 at $q\sim0.95$.

\subsubsection{Mock binary star experiments} \label{subsubsubsec:mock_bianry_9000}

\begin{figure*}
\centering
\includegraphics[trim=0 0 2 0, clip, width = 15cm]{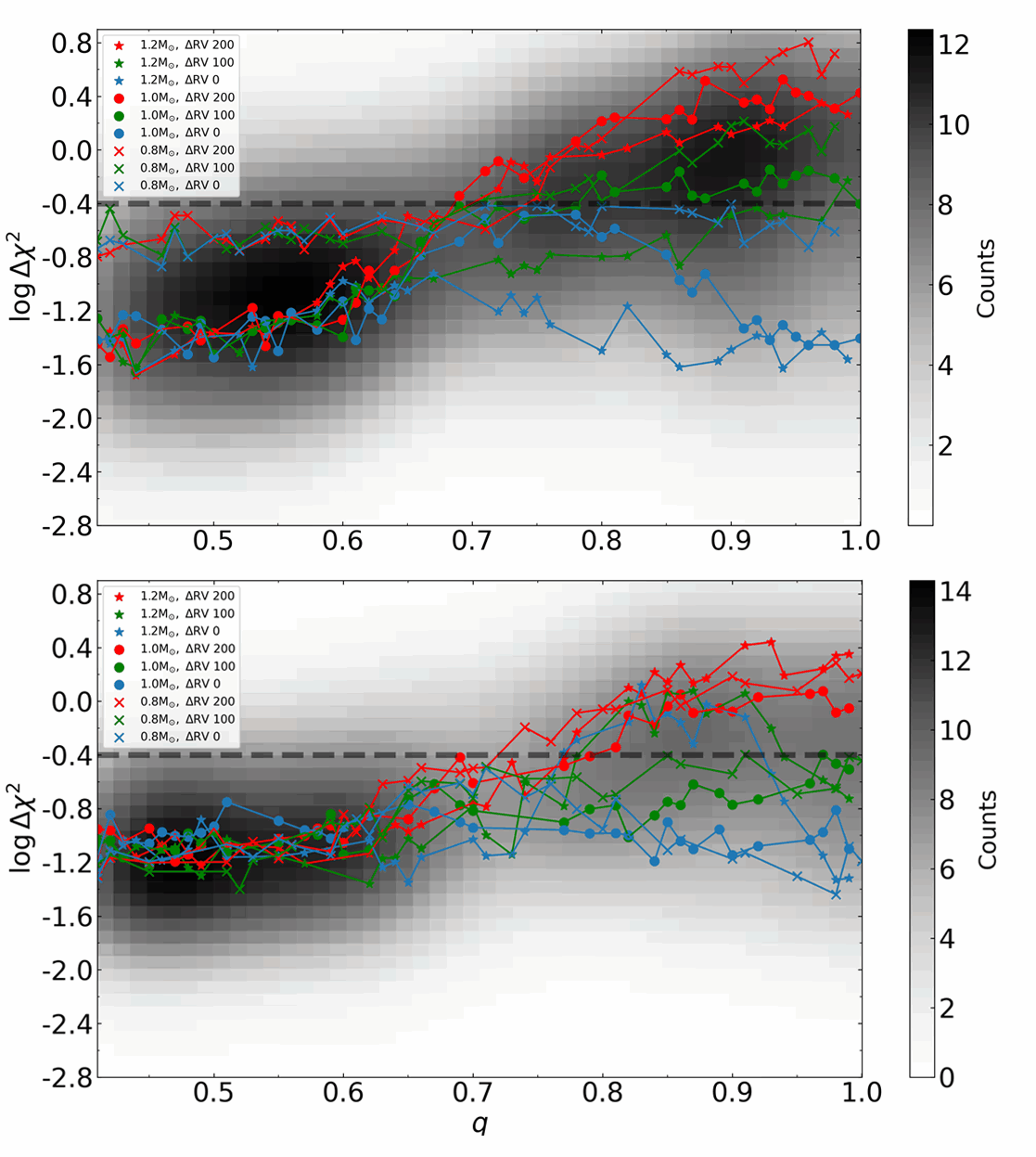}
\caption{The $q$ versus $\log \Delta \chi^{2}$ distributions in mock binary experiments $E1$ and $E2$. The background is the density distribution before averaging the data with different signal-to-noise ratios. The dashed lines are at $\log \Delta \chi^{2}>-0.4$.}
\label{fig:E1_E2}
\end{figure*}

As listed in Table~\ref{table:get_criteria_value}, we performed two experiments ($E1$ and $E2$) to determine a certain $\Delta \chi^{2}$ value that guides our models to identify binary stars in different parameter spaces and SNRs. The experiment $E1$ is designed for solar-like stars, while $E2$ is focused on objects with relatively low [M/H]. For $E1$, we selected one value in each parameter, so there are 9 unique combinations of parameters [M/H], [$\alpha$/M], $\rm log \tau$, mass, and $\Delta$RV. Then, for each combination, we set up 50 identical copies and assigned $q$ values randomly following a uniform distribution of $\mathcal{U}$(0.4, 1.0). We called these unique 450 binaries (9 $\times$ 50) as a collection. Finally, this collection was replicated for 20 sets, and they followed the SNR distribution of LAMOST DR9. Following the method in Section~\ref{subsec:mock_binary}, we obtained 9000 synthetic ‘semi-empirical’ spectra while the experiment $E2$ follows the same procedures. Adopting the single-star and binary-star models fitting in Section~\ref{sec:Single-star-model} and \ref{sec:Binary-star-model}, we calculated the $\Delta \chi^{2}$ for $E1$ and $E2$. One should note that we adopted the average value of $\Delta \chi^{2}$ for each unique binary with different SNRs to consider the influence of SNRs, so there are 450 points plotted in Figure~\ref{fig:E1_E2} along the $q$. 

As shown in Figure~\ref{fig:E1_E2}, for sources with $q \gtrsim 0.7$, $\log(\Delta \chi^{2})$ is a strong function with RV offset, where the larger radial velocity difference, the larger $\log(\Delta \chi^{2})$ value. Because in high mass ratio systems, the spectral contributions of both components are significant, and as the RV offset increases, the split in the absorption lines becomes more pronounced. For sources with $q\lesssim0.7$, the lower spectral contribution of the secondary star is submerged in the spectral contribution of the primary star, resulting in sources with different RV offsets being mixed and indistinguishable. At this point, $\log(\Delta \chi^{2})$ is all lower than -0.4, i.e., $\Delta \chi^{2}<0.4$. Therefore, $\Delta \chi^{2}>0.4$ can serve as the threshold for binary identification in this work. Moreover, under this threshold, both $E1$ and $E2$ indicate that our method can identify binaries with RV offsets > 100~km$\,$s$^{-1}$. We can conclude that our model has good identification for systems with a mass ratio greater than 0.7 and RV offsets greater than 100 km$\,$s$^{-1}$.

\subsection{Radial velocity reversal} \label{subsec:reverse_RV}

The binary-star model fitting for high mass-ratio sources suffers challenges in accurately attributing RVs to their respective components. This difficulty arises from the similarity in spectra between the two components in such systems and the blending of flux at low spectral resolutions. While completely resolving this issue is difficult, we presented a strategy, RV reversal, to mitigate it as much as possible. We rank $\Delta\chi^{2}_{\rm e}$, the differences between single-star and binary-star model fits in each epoch, from largest to smallest. Following this sequence, we proceeded by sequentially adding each spectrum to the binary-star model fitting queue, one by one. This sequence is based on the rationale that the spectrum exhibiting the highest $\Delta \chi^{2}_{\rm e}$ is likely to contain the most prominent features resulting from RV offsets between the components. In each reanalysis, we applied an RV reversal process. This requires interchanging the previously fitted RVs attributed to components 1 and 2. Subsequently, we retained the configuration with the lowest $\chi^{2}_{\rm b}$ value in each epoch (denoted as $\chi^{2}_{\rm b,e}$), which serves as the foundation for the subsequent fitting queue. Throughout this process, we compared the $\chi^{2}_{\rm b,e}$ values before and after the RV reversal and define the difference as $\delta \chi^{2}_{\rm b,e}$, which can be used to judge the significance of the RV reversal. As we retained the minimum $\chi^{2}_{\rm b,e}$, $\delta \chi^{2}_{\rm b,e}$ is always $\leq$ 0. Among the 4,848 binary candidates we released (Section~\ref{subsec:SB2_candidates}), 593 sources with multiple observations and 351 sources having a single observation achieved better fitting results through RV reversal.

\begin{figure*}
\centering
\includegraphics[width = 18cm]{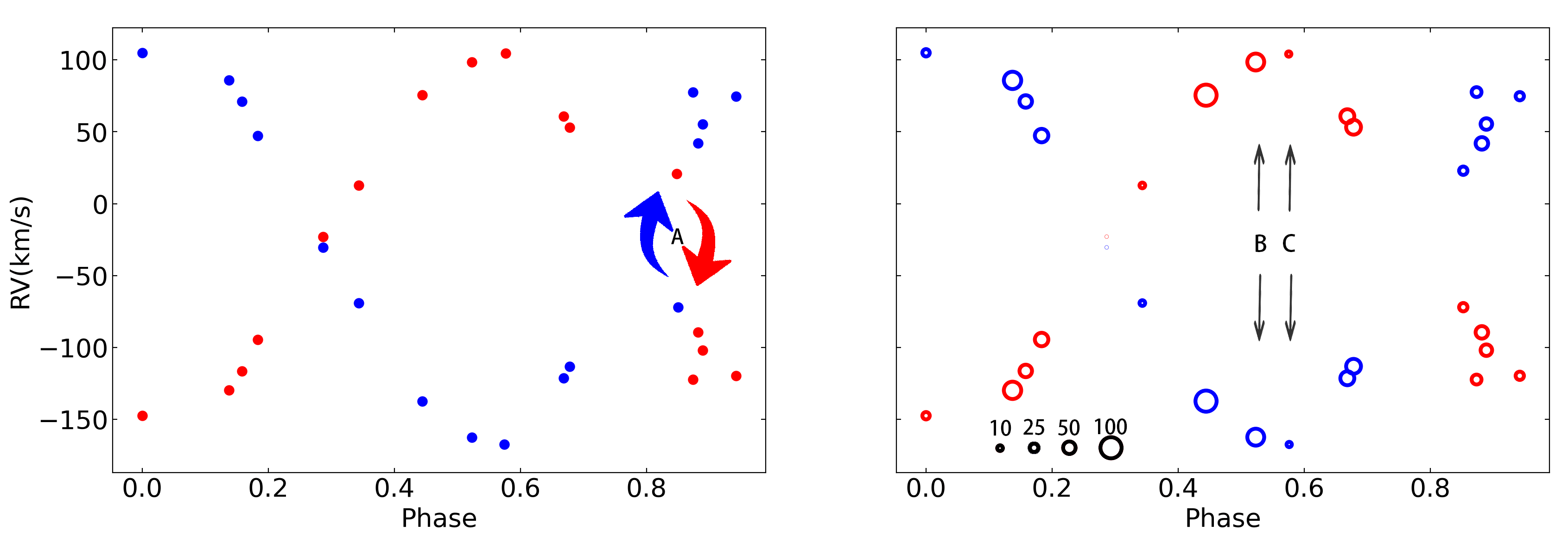}
\caption{Comparing before (left panel) and after (right panel) the radial velocity reversal for ASAS-SN J064726.41+223431.7. The red and blue dots (or circles) represent the radial velocities of the primary and secondary stars, respectively. Left panel: The point pair `$\tt A$' indicates the one-pair radial velocities that were incorrectly assigned to the component stars, and the blue and red arrows illustrate a radial velocity reversal. Right panel: The size of the circles in the right panel represents 100 times $|\delta \chi^{2}_{\rm b,e}|$.}
\label{fig:reverse_RV}
\end{figure*}

As illustrated in Figure~\ref{fig:reverse_RV}, we phased the radial velocity (RV) data of the binary candidate, identified as ASAS-SN J064726.41+223431.7, based on the period determined from the All-Sky Automated Survey for Supernovae (ASAS-SN). The RV pair `$\tt A$' in the left panel represents incorrectly assigned radial velocities. After the RV reversal, the RVs of this pair have been reversed correctly in the right panel. As expected, in this binary system with a $q$ close to 1, the larger the difference in RVs between the two components, the larger the absolute value (size) of $\delta \chi^{2}_{\rm b,e}$ and the RVs of components are also less prone to being inaccurately assigned. In addition, the signal-to-noise ratio of spectra also has a negative correlation with the value of $|\delta \chi^{2}_{\rm b,e}|$. For example, there are two RV pairs `$\tt B$' and `$\tt C$' in the right panel with almost the same $\Delta \rm RV$ at phases 0.523 and 0.575, but $\delta \chi^{2}_{\rm b,e}$ is indeed different. This discrepancy arises due to the different signal-to-noise ratios: point `$\tt B$' has a signal-to-noise ratio of 165.78, while point `$\tt C$' has a signal-to-noise ratio of 37.36. $\delta \chi^{2}_{\rm b,e}$ was added to the released data (Table~\ref{table:4848 binary candidates}) to represent the difference in the spectral fitting of each epoch before and after RV reversal. It's important to note that, after comparing the spectral periods with the ASAS-SN light curve periods (in Section~\ref{subsec:Joker}), we estimate that approximately 40\% of the sources (with observational times $\geq7$) have their RVs accurately assigned to the correct component star for each epoch after the correction of RV reversal. 

\subsection{Model fitting and uncertainty} \label{subsec:MCMC}

Combining spectral and photometric data (i.e., $Gaia$ DR3 and 2MASS), astrometric data ($Gaia$ DR3 parallax), and Galactic extinction ($\tt dustmaps$; \citep{2019ApJ...887...93G}), we utilize Markov Chain Monte Carlo (MCMC) fitting implemented through the $\tt emcee$ package \citep{2013PASP..125..306F} to fit and constrain the stellar parameters, distances, and extinction.
The $\tt emcee$ package uses an ensemble sampler to explore the parameter space of a given probability distribution efficiently and uses multiple chains in parallel to explore the parameter space. It can efficiently find the global maximum or minimum of the distribution and also provides a better estimate of the uncertainties associated with the inferred parameters. In the simulation, the construction of likelihood estimation is based on: the binary-star model, which can predict the spectrum; the MIST model can predict the absolute magnitudes vector $\overrightarrow{M_{\rm \lambda}}$, including the $J$, $H$, and $Ks$ bands luminosity of 2MASS, and $G$, $G_{\rm BP}$, and $G_{\rm RP}$ bands luminosity of $Gaia$ DR3, which, combined with the observed magnitudes, can be combined to estimate model distance and extinction.

\begin{figure*}
\centering
\includegraphics[width = 15cm]{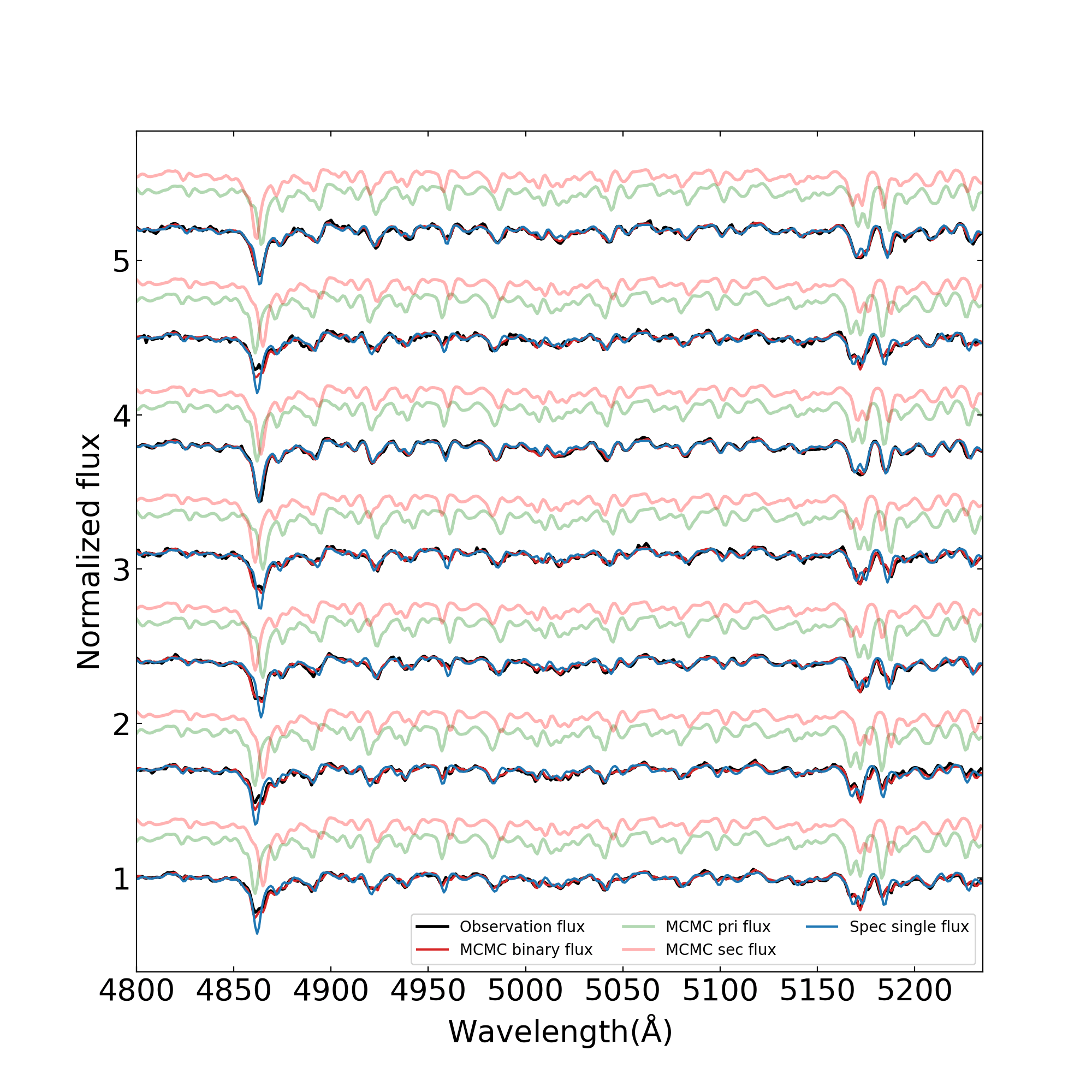}
\caption{The spectral fitting results in different epochs for J0913. The black curve represents the observed spectra. The deep red curves represent the binary-star spectra generated using the median spectral parameters derived from the posterior distributions of the MCMC fitting, while the deep blue curves show the results of single-star model fitting. The light green and red curves represent the primary and secondary star spectra.}
\label{fig:Multi_epoch_object}
\end{figure*}

\begin{figure*}
\centering
\includegraphics[trim= 35 10 0 0, width = 19cm]{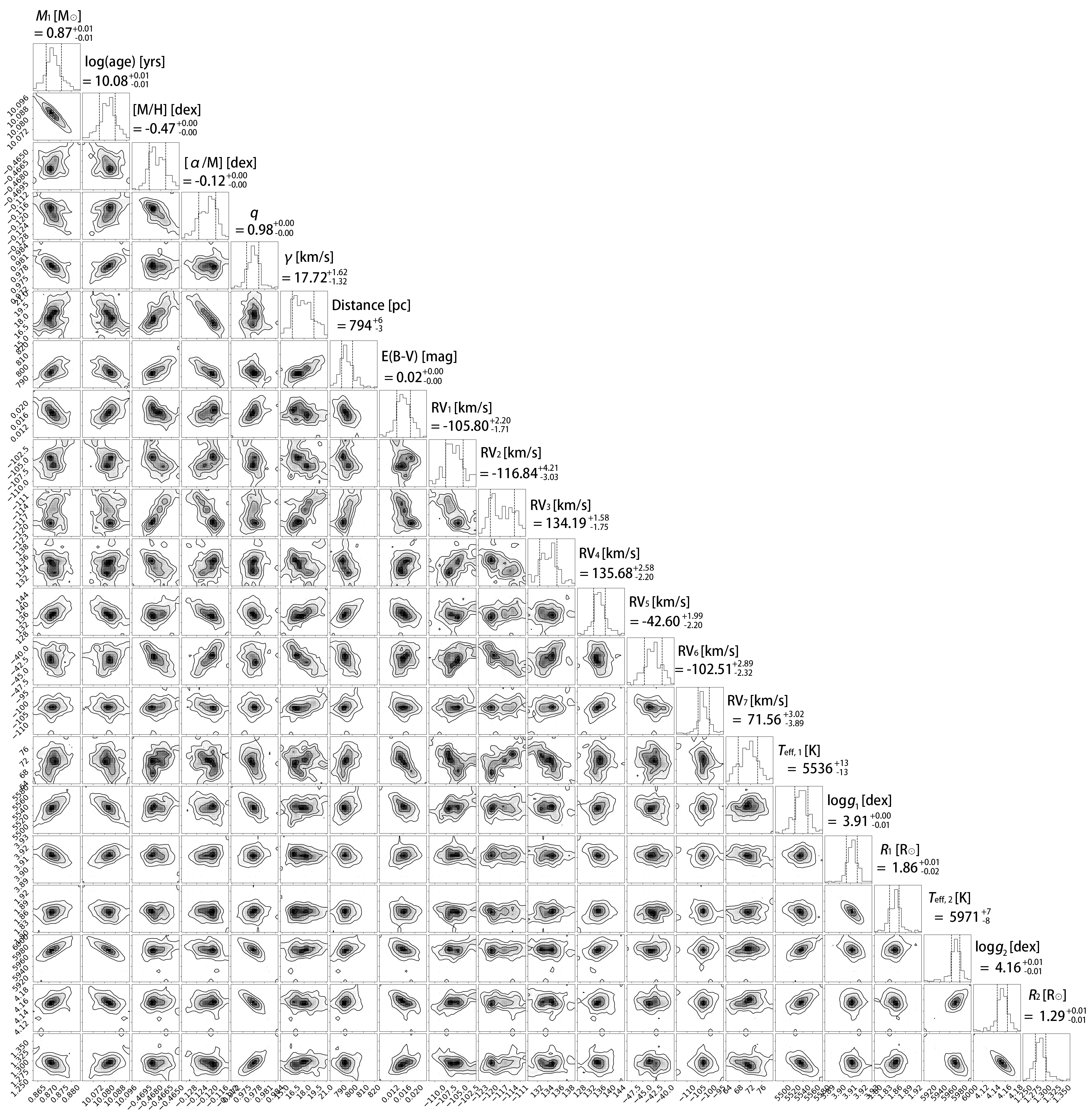}
\caption{The marginalized posterior probability distributions from the MCMC fitting for J0913 with 7 observational epochs (with RV values for the primary component). The parameter values are the median of the one-dimensional distributions, while the vertical dashed lines are located at the 16th and 84th percentiles.}
\label{fig:corner_plot}
\end{figure*}

The likelihood function is 
\begin{equation}\label{equ:likelihood_function}
\begin{split}
\ln p&(\overrightarrow{\theta}_{b}|\overrightarrow{f}, \overrightarrow{m}_{\rm \lambda}, E(B-V), \varpi) = \\
&-\frac{1}{2}\times \left\{ \left[ \frac{(\overrightarrow{f}_{\rm fit}-\overrightarrow{f}_{\rm obs})^{2}}{\overrightarrow{\sigma}_{\rm f}^{2}} + \ln(2\overrightarrow{\sigma}_{\rm f}^{2}) \right] \right .\\
&+\left[\frac{(\overrightarrow{m}_{\rm \lambda, fit}-\overrightarrow{m}_{\rm \lambda, obs})^{2}}{\overrightarrow{\sigma}_{\rm \lambda}^{2}} + \ln(2\overrightarrow{\sigma}_{\rm \lambda}^{2})\right]\\
&+\left[\frac{(E(B-V)_{\rm fit}-\textit{E}(B-V)_{\rm dp})^{2}}{\sigma_{\rm ebv}^{2}}\right]\\
& \left. + \left[\frac{(\varpi_{\rm fit}-\varpi_{\rm obs})^{2}}{\sigma_{\varpi}^{2}} + \ln(2\sigma_{\varpi}^{2}) \right] \right\} + \rm const, 
\end{split}
\end{equation}
where $\overrightarrow{m}_{\rm \lambda, fit}$ and $\overrightarrow{m}_{\rm \lambda, obs}$ are the fitting and observational magnitudes in different bands, i.e., $J$, $H$, $Ks$, $G$, $G_{\rm BP}$ and $G_{\rm RP}$ bands; $\overrightarrow{f}_{\rm fit}$ and $\overrightarrow{f}_{\rm obs}$ are the fitting and observational normalized spectral flux with wavelength ranging from 3950 to 5750 \AA; $E(B-V)_{\rm fit}$ and $E(B-V)_{\rm dp}$ are the fitting and $\tt dustmaps$ extinction; $\varpi_{\rm fit}$ and $\varpi_{\rm obs}$ are the fitting and $Gaia$ DR3 parallax; $\overrightarrow{\sigma}_{f}$, $\overrightarrow{\sigma}_{\rm \lambda}$ and $\sigma_{\varpi}$ are the observational errors of the above-mentioned spectral flux, photometric magnitude and parallax, respectively, while $\sigma_{\rm ebv}$ is set to a typical value 0.01. As listed in Table~\ref{table:4848 binary candidates}, we adopt the median values of the marginalized probability distributions fitted by MCMC as stellar parameters. Meanwhile, the calculation of the uncertainties is based on the corresponding values in the 16th and 84th percentiles. The spectral fitting results and the marginalized probabilities for a multi-epochs binary candidate with ASAS-SN name J091352.13+025734.8 (hereafter J0913) are shown in Figure~\ref{fig:Multi_epoch_object} and \ref{fig:corner_plot}, respectively, which demonstrate that our model can fit the spectra well in every phase, and the MCMC method effectively constrains various parameters.

\section{Method Performance}\label{sec:Detection_Efficiency}

\subsection{RV standard star fitting} \label{subsubsubsec:RV_standard_star}

Besides the model performances discussed in the mock binary fitting (Section~\ref{subssec:Find_the_criteria}), we utilized the RV standard stars to evaluate the method and certify the criterion for binary identification by applying the single-star and binary-star models fitting. The RV standard stars can be treated as single stars that maintain a stable RV over a long time baseline. \citet{2018AJ....156...90H} assembled a catalog of 18080 RV standard stars (3~$\sigma_{\rm RV}<$~240 m/s) which are observed by APOGEE survey at least 3 times covering time baseline more than 200 days. We cross-matched them with the LAMOST LRS DR9 catalog in a radius of 3$"$. This process led to 3239 sources in total, including 934 objects with $\log g>3.5$. For these 934 stars, we both adopted the single-star and binary-star models to fit their spectra and found 43 objects with $\Delta \chi^{2} = (\chi^{2}_{s} - \chi^{2}_{b}) > 0$ which means that $\sim95\%$ of the RV standard stars preferred single star model when using $\Delta \chi^{2} > 0.0$. When we set $\Delta \chi^{2} > 0.4$ to identify the binary candidates, there is no object remaining, which indicates that this criterion is reliable for binary selection. 

\begin{figure*}
\centering
\includegraphics[width = 18cm]{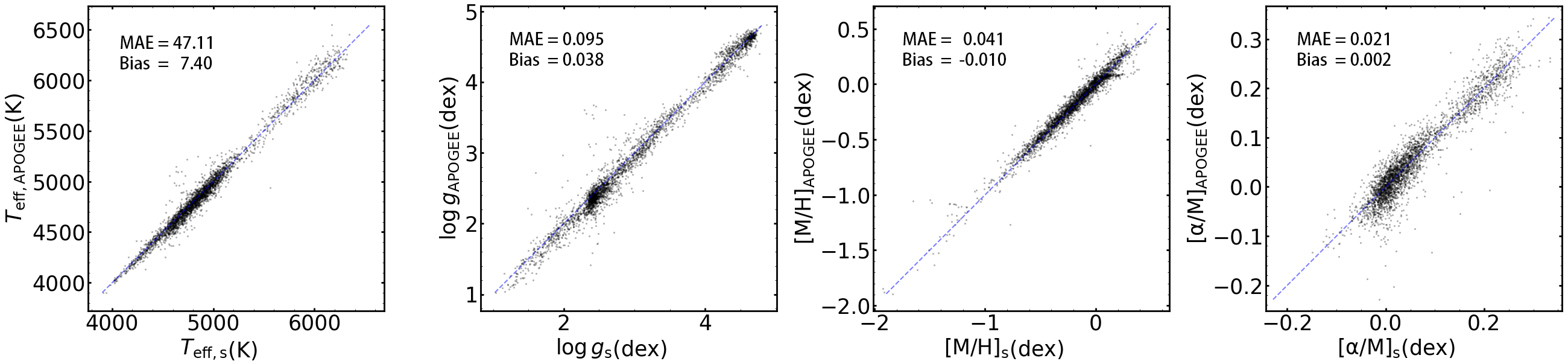}
\caption{The comparison for the spectral parameters of RV standard stars between single-star model fitting and APOGEE catalog. The MAE and bias have the same definitions as Figure~\ref{fig:one-to-one_plot}.}
\label{fig:RV_standard}
\end{figure*}

The single-star model fitting results versus the parameters released from APOGEE of the whole 3239 objects are shown in Figure~\ref{fig:RV_standard}. In each panel, we provided MAE and bias values for the comparison distributions, showcasing tightly distributed points along the one-to-one lines with small scatters. This illustrates the reliability of our predictions for the parameters of both main-sequence and giant stars, despite our primary focus on the former.

\subsection{Mock single star spectral fitting} \label{subsubsec:Mock_single-star_spectra}

The false positive rate (FPR) is an important metric for evaluating machine learning models, requiring us to evaluate the probability of single-star samples being classified as binary stars.
We synthesized 10,000 single-star spectra employing the stellar parameters ranges identical to the primary stars in Table~\ref{table:Test_mock_binary_parameters} and the same spectral synthesis method as described in Section~\ref{sec:Single-star-model}. After applying single-star and binary-star models fitting to these spectra, we found that 234 single stars (out of 10,000) were classified as binary stars under the threshold of $\Delta \chi^{2} > 0.4$, which implies that the FPR of our model is 2.3\%.

 \begin{table*}
   \caption{The summary of the parameters of the mock binary stars in experiments $E1$ and $E2$.}
     \label{table:get_criteria_value}
   \begin{center}
   \begin{tabular}{ccccccc}\hline \hline
 Experiment name & [M/H] & [$\alpha$/M] & $\log \tau$ & Mass & $\Delta RV$ & q \\
 & dex & dex  &  years & $M_{\odot}$&  km~s$^{-1}$ &  \\
\hline
 $E1$ & 0.0 & 0.0 & 9.65 & $\mathcal{T}$(0.8, 1.0, 1.2) & $\mathcal{T}$(0, 100, 200) & $\mathcal{U}$(0.4, 1.0) \\
 $E2$ & -1.0 & 0.4 & 9.65 & $\mathcal{T}$(0.8, 1.0, 1.2) & $\mathcal{T}$(0, 100, 200) & $\mathcal{U}$(0.4, 1.0) \\
\hline\noalign{\smallskip}
  \end{tabular}
  \begin{tablenotes}
  \item[1](Notes: $\mathcal{U}$($a$, $b$) represent a uniform distribution from $a$ to $b$. $\mathcal{T}$(e, f, g) indicates one value is selected to combine with other parameters in turn.)
  \end{tablenotes}
  \end{center}
\end{table*}

\subsection{Confusion Matrix} \label{subsec:Confusion_Matrix}

Evaluating the performance of spectral model fitting based on machine learning typically uses metrics like accuracy, precision, and recall. These metrics are derived from the confusion matrix, which includes values for TP (True Positives), FP (False Positives), TN (True Negatives), and FN (False Negatives). In this study, TP and FN correspond to positive samples (i.e., SB2s) that are correctly and incorrectly predicted, while TN and FP represent cases involving negative samples (i.e., single stars). Their definitions are as follows:
\begin{equation}\label{equ:Accuracy}
    \rm Accuracy = \frac{TP + TN}{TP + FP + FN + TN},
\end{equation}
\begin{equation}\label{equ:Precision}
    \rm Precision = \frac{TP}{TP + FP},
\end{equation}
\begin{equation}\label{equ:Precision2}
    \rm Recall = \frac{TP}{TP + FN},
\end{equation}

\begin{figure}
\centering
\includegraphics[width = 6cm]{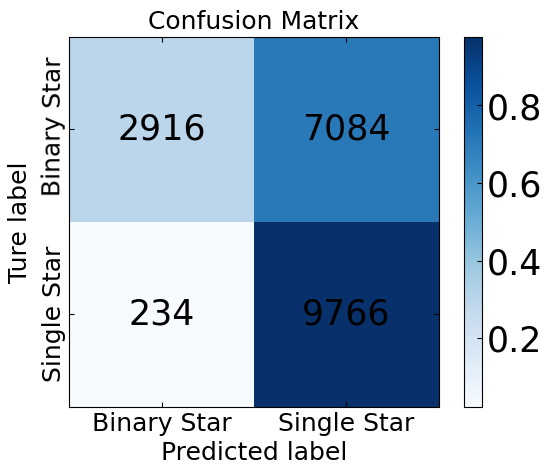}
\caption{Confusion matrix of the model on the testing set. In the order of top to bottom, left to right, they represent TP, FP, FN, and TN, respectively.}
\label{fig:Confusion_Matrix}
\end{figure}

Based on the single-star and binary-star models fittings for mock binary-star (in Section~\ref{subsec:mock_binary}) and single-star spectra (in Section~\ref{subsubsec:Mock_single-star_spectra}), the confusion matrix results are illustrated in Figure~\ref{fig:Confusion_Matrix}. Under the condition of an equal ratio of mock single-star and binary-star spectra, the values of accuracy, precision, and recall in this work are 63.4\%, 92.6\%, and 23.0\%, respectively. For this study, the most important metric is precision, which measures the reliability of our model in identifying SB2s. The precision value (92.6\%) indicates that the sample of SB2s selected using our method would only have a few percent of false positive contamination.

\section{Results and Cross Validation} \label{sec:Result&Validation}

\subsection{SB2 candidates and the parameters} \label{subsec:SB2_candidates}

We employed single-star and binary-star models in the fitting of 3,584,214 LAMOST LRS DR9 spectra. After applying the experimental criterion of $\Delta \chi ^{2} > 0.4$, RV reversal, MCMC fitting, and visual screening, we obtained a total of 4848 binary candidates with stellar parameters in the range of 4000~K$\lesssim T_{\rm eff}\lesssim$7000~K, 3.5$\lesssim \log g \lesssim $5.0, -2.0$\lesssim$[M/H]$\lesssim$0.5 and -0.5$\lesssim[\alpha / \rm M]\lesssim$0.5. These SB2 candidates constitute 0.14\% (4,848/3,584,214) of the total spectra, and their stellar parameters are listed in Table~\ref{table:4848 binary candidates}. These parameters, particularly temperature, have a distribution range narrower than that of the training samples. This is done to enhance the accuracy of the released data, eliminating potential inaccuracies in parameter predictions at the boundaries of the training parameters. 

\begin{figure*}
\centering
\includegraphics[trim=7 0 20 0, clip, width = 18.5cm]{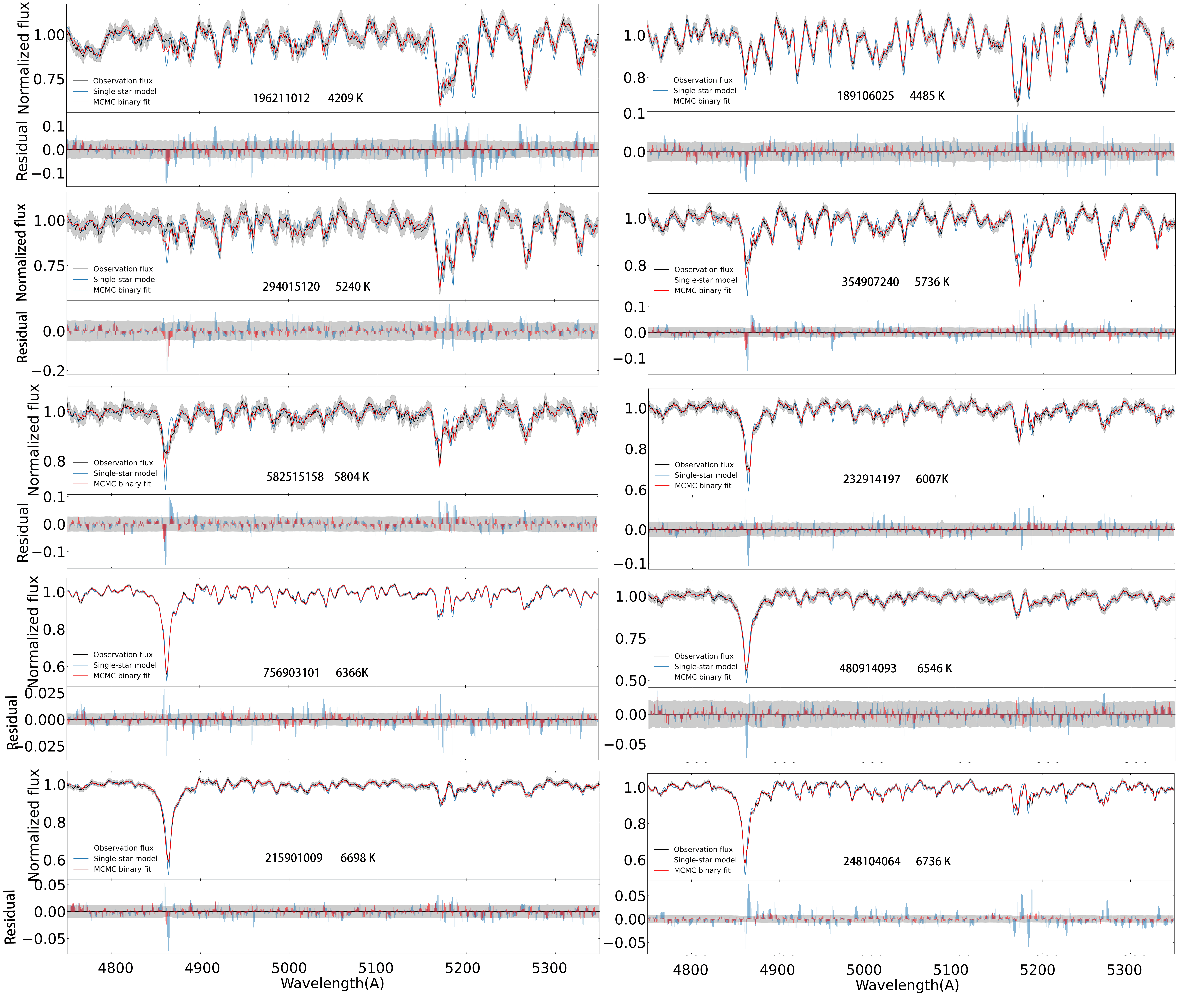}
\caption{Comparison of single-star and binary-star model fittings for binary candidates labeled with LAMOST $Obsid$.}
\label{fig:ASAS_samples}
\end{figure*}

The comparison of single-star and binary-star model fittings for 10 randomly selected binary samples, sorted by the temperature of the primary star, is shown in Figure~\ref{fig:ASAS_samples}. The features of their absorption lines, such as $\rm H \beta$ and Mg I triplet, are well-described by the binary star model fitting, exhibiting clear differences from the results of the single-star model.

\begin{figure}
\centering
\includegraphics[trim=7 0 0 0, clip, width = 15cm]{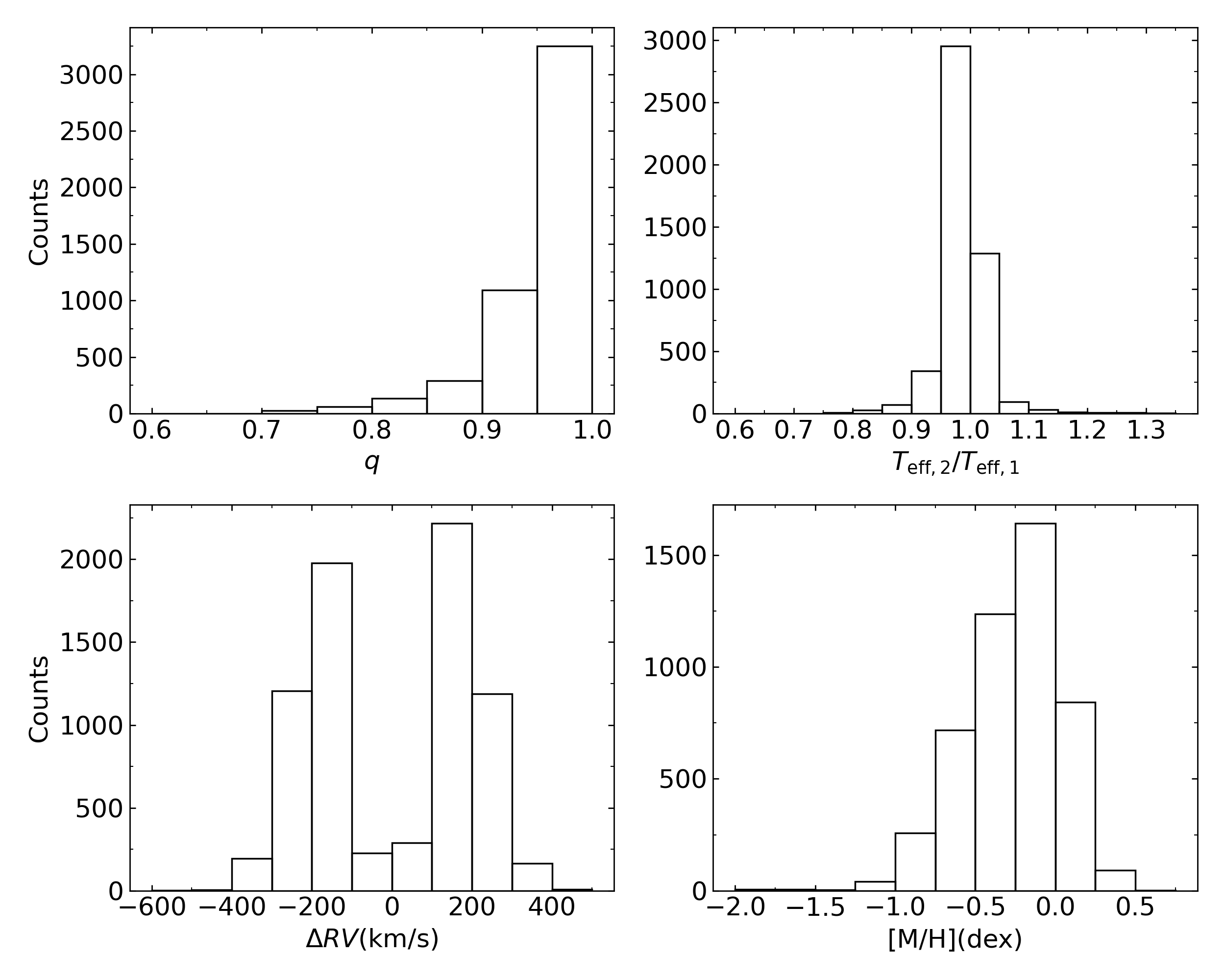}
\caption{Distribution plots of mass ratio $q$, temperature ratio $T_{\rm eff, 2}/T_{\rm eff, 1}$, radial velocity difference $\Delta \rm RV$ ($\rm RV_{1}-RV_{2}$), and metallicity [M/H].}
\label{fig:Parameters_distributions}
\end{figure}

The distributions of mass ratio $q$, temperature ratio $T_{\rm eff, 2}/T_{\rm eff, 1}$, radial velocity difference $\Delta \rm RV$ ($=\rm RV_{1}-RV_{2}$), and metallicity [M/H] are shown in Figure~\ref{fig:Parameters_distributions}. For the mass ratio, most values are greater than 0.7, with 89.6\% (4343 objects) of the sources having a mass ratio greater than 0.9. The temperature ratio $T_{\rm eff, 2}/T_{\rm eff, 1}$ between the secondary and primary stars is concentrated between 0.90 and 1.05, with the peak occurring near 1.00. This is consistent with the distribution of the mass ratio, indicating that the two components of most of binary candidates have similar masses and temperatures. For $\Delta \rm RV$, it exhibits a symmetric distribution with respect to zero, with concentrations in the ranges of $-300$ to $-100$ km~s$^{-1}$ and 100 to 300 km~s$^{-1}$. This suggests that our method is more sensitive to sources with significant radial velocity differences, which is consistent with the results of the mock binary spectra experiments discussed in Section~\ref{subssec:Find_the_criteria}. Additionally, the metallicity of the binary star candidates is primarily distributed between $-0.75$~dex and 0.25~dex, with a peak near $-0.125$~dex.

\begin{table*}
   \caption{List of the parameters of 4848 binary candidates.}
     \label{table:4848 binary candidates}
   \begin{center}
   \begin{tabular}{llllcl}\hline \hline
 Index & Quantity & Column & Format & Units & Description\\
\hline
1 & Obsid  & Obsid   & List(n,) &     ---        & LAMOST observational ID (unique for each fits file; n is n epochs)\\
2 & BJDmid & BJDmid    & List(n,)  &    days        & Barycentric Julian Date of the middle of exposure\\
3 &  R.A.    &  R.A.    & Float  &    degrees     & Right ascension in decimal degrees (J2000)\\
4 &  DEC.   &  DEC.    & Float  &    degrees     & Declination in decimal degrees (J2000)\\
5 &  SNR$_{\rm g}$ &  SNR$\_\rm g$    & List(n,)   &    ---         & SNR ratio for the $g-$band \\ \hline
6 &  $M_{1}$    &  M\_1  & Float   &    M$_{\odot}$ & Mass of the primary component\\
7 &  $M_{\rm 1,err}$ &  M\_1\_err & Float &  M$_{\odot}$ & Typical error of $M_{1}$  \\
8 &  $q$     &  q     & Float   &    ---         & Mass ratio $q$ = $M_{2}/M_{1}$ \\
9 &  $q_{\rm err}$ &  q\_err    &  Float & ---        & Typical error of $q$\\
10 & [M/H]  & MH   &  Float  &   dex          & Metallicity \\
11 & [M/H]$_{\rm err}$ & MH\_err & Float & dex         & Typical error of [M/H]\\
12 & [$\rm \alpha$/M] & AlphaM & Float   &  dex           & Alpha elements Metallicity\\
13 & [$\rm \alpha$/M]$_{\rm err}$ & AlphaM\_err & Float & dex         &Typical error of [$\rm \alpha$/M] \\
14 & $T_{\rm eff,1}$ & Teff\_1  & Float    &  K             & Effective temperature of primary component \\
15 & $T_{\rm eff,1,err}$ & Teff\_1\_err & Float & K         & Typical error of $T_{\rm eff,1}$  \\ 
16 & $\log g_{1}$  & logg\_1     & Float     & dex       &  Surface gravity of primary component \\
17 & $\log g_{1,err}$ & logg\_1\_err & Float     & dex       & Typical error of $\log g_{1}$  \\
18 & $R_{1}$  & R\_1     & Float     & R$_{\odot}$       &  Radius of primary component \\
19 & $R_{1,err}$ & R\_1\_err & Float     & R$_{\odot}$       & Typical error of $R_{1}$ \\
20 & $T_{\rm eff,2}$ & Teff\_2  & Float    &  K      & Effective temperature of secondary component \\
21 & $T_{\rm eff,2,err}$ & Teff\_2\_err & Float & K         & Typical error of $T_{\rm eff,2}$\\
22 & $\log g_{2}$  & logg\_2    & Float     & dex       &  Surface gravity of secondary component \\
23 & $\log g_{2,err}$ & logg\_2\_err & Float     & dex       & Typical error of $\log g_{2}$\\
24 & $R_{2}$   & R\_2      & Float     & R$_{\odot}$       &  Radius of primary component \\
25 & $R_{2,err}$ & R\_2\_err & Float     & R$_{\odot}$       & Typical error of $R_{2}$\\
26 & $RV_{1}$    & RV\_1     & List(n,) & km$\,$s$^{-1}$      & Radial velocity of primary component \\ 
27 & $RV_{\rm 1, err}$  & RV\_1\_err    & List(n,) & km~s$^{-1}$      &Typical error of $RV_{1}$ \\ 
28 & $RV_{2}$    & RV\_2     & List(n,) & km$\,$s$^{-1}$      & Radial velocity of secondary component \\ 
29 & $RV_{\rm 2, err}$  & RV\_2\_err    & List(n,) & km~s$^{-1}$      &Typical error of $RV_{2}$ \\ 
30 & $\gamma$   & gamma     & Float     & km$\,$s$^{-1}$      &   Centre-of-mass heliocentric velocity \\
31 & $\gamma_{\rm err}$  & gamma\_err   & Float     & km$\,$s$^{-1}$      & Typical error of $\gamma$ \\
32 & Dist   & Distance     & Float     & pc        &  Distance \\
33 & Dist$_{\rm err}$ & Distance\_err  & Float & pc        & Typical error of Distance\\
34 & E(B-V)    & ebv     & Float     & mag       & Extinction \\
35 & E(B-V)$_{\rm err}$  & ebv\_err       & Float     & mag       & Typical error of Extinction\\
36 & log(age)  & logage     & Float     & yr      & Age \\
37 & log(age)$_{\rm err}$ & logage\_err & Float     & yr      & Typical error of Age \\
38 & $\Delta\chi^{2}_{\rm e}$ & Delta\_chi2\_e       & List(n,)    &       &  Difference between single-star and binary-star models fitting spectra 
\\
   &                          &                      &              &        & in each epoch \\
39 & $\delta \chi^{2}_{\rm b,e}$ & delta\_chi2\_be & List(n,) &       & Difference between the $\chi^{2}_{\rm b,e}$ values before and after the RV reversal,   \\
   &             &         &    &       & ($\chi^{2}_{\rm b,e}$: difference between the binary-star model and observational spectra)   \\
\hline\hline\noalign{\smallskip}
  \end{tabular}
    \begin{tablenotes}
  \item[1] This table is available in its entirety in machine-readable form.
  \end{tablenotes}
  \end{center}
\end{table*}

\subsection{Color-Magnitude Diagram} \label{subsec:CMD}

An important way to verify the authenticity of our identified spectral binary candidates is to examine their distribution on the color-magnitude diagram (CMD), which is mainly based on the fact that for the same color, the position of the binary star is usually above the single star main sequence \citep{2013MNRAS.436.1497L, 2018MNRAS.476..528E, 2022ApJS..258...26Z}. This phenomenon is primarily attributed to the luminosity contribution of the secondary star \citep{2018MNRAS.476..528E}. 

To construct a CMD, distance, and extinction for each object are essential for the accurate calculation of the absolute magnitude. We assigned distances measured by $Gaia$ DR3 to each source. Extinction correction was again based on $\tt dustmaps$. Assuming a canonical interstellar reddening law (i.e., $R_{\rm V} = 3.1$, \citealt{1989ApJ...345..245C}), we corrected the observed bands from $Gaia$ and 2MASS using the 
relative extinction values of $A_{\rm \lambda}/A_{\rm V}$ = 0.7889, 1.0139, and 0.5965 for the $G$, $G_{\rm BP}$, and $G_{\rm RP}$ bands \citep{2019ApJ...877..116W}, while $A_{\rm \lambda}/A_{\rm V}$ = 0.2557, 0.1513, and 0.0976 for the $J$, $H$, and $Ks$ bands, respectively \citep{2019ApJ...887...93G}.

\begin{figure*}
\centering
\includegraphics[width = 18cm]{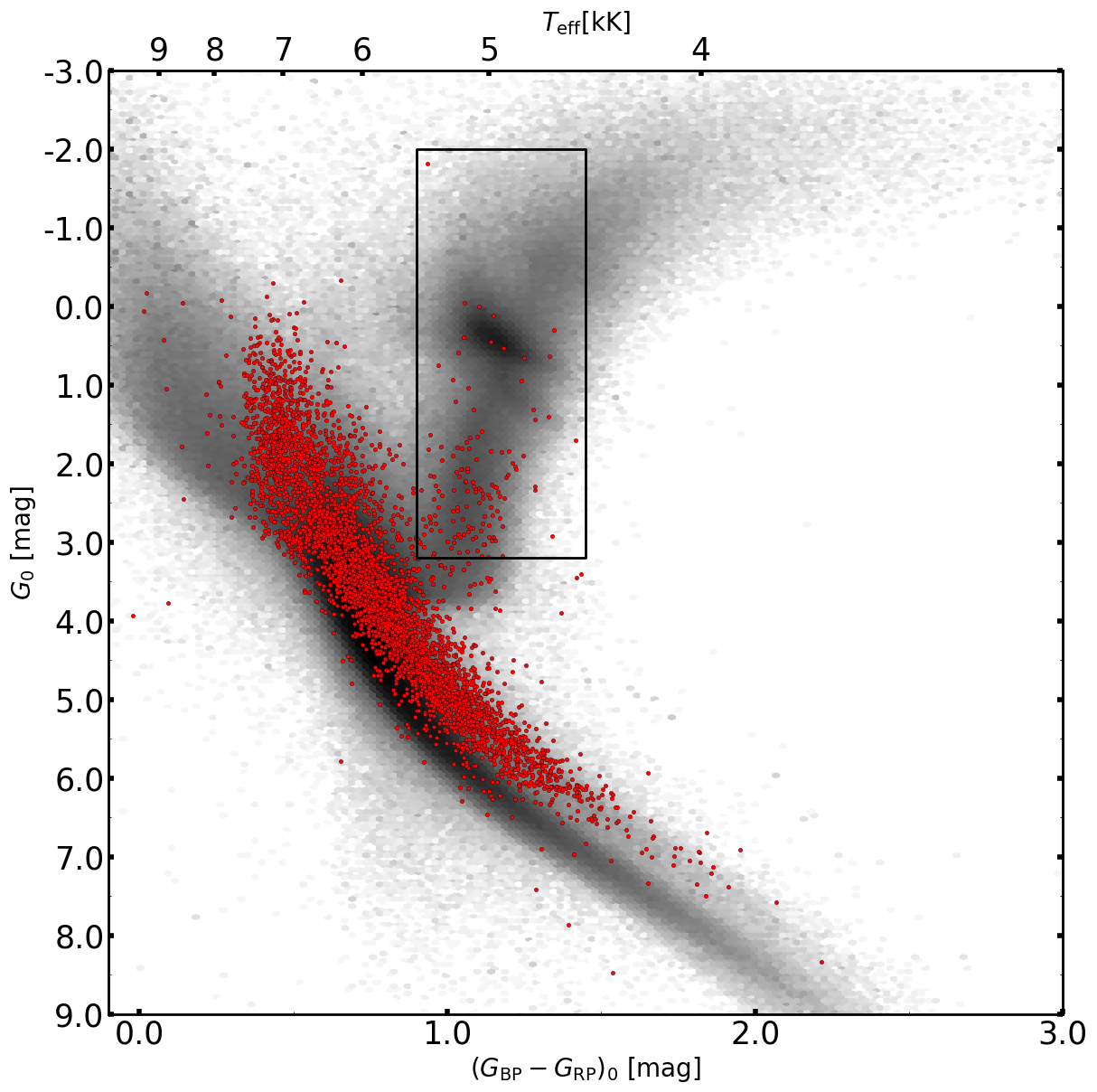}
\caption{The CMD (red points) of the SB2 candidates found by this work. The black hot map shows the distribution of all LAMOST LRS DR9 objects after extinction correction. The black box indicates the region having binary candidates that include giant stars. }
\label{fig:CMD}
\end{figure*}

Based on the $Gaia$ DR3 photometry, we plot the CMD of $(G_{\rm BP}-G_{\rm BP})_{0}$ versus $G_{0}$ for the binary candidates (red points) in Figure~\ref{fig:CMD} where the black hot map background is all sources from LAMOST LRS DR9 after the same extinction correction as mentioned above. It is clear that the locus of the CMD of binary candidates falls tightly above the main sequence, which is a robust test for our binary candidates.

\begin{figure*}
\centering
\includegraphics[trim=7 0 20 0, clip, width = 18cm]{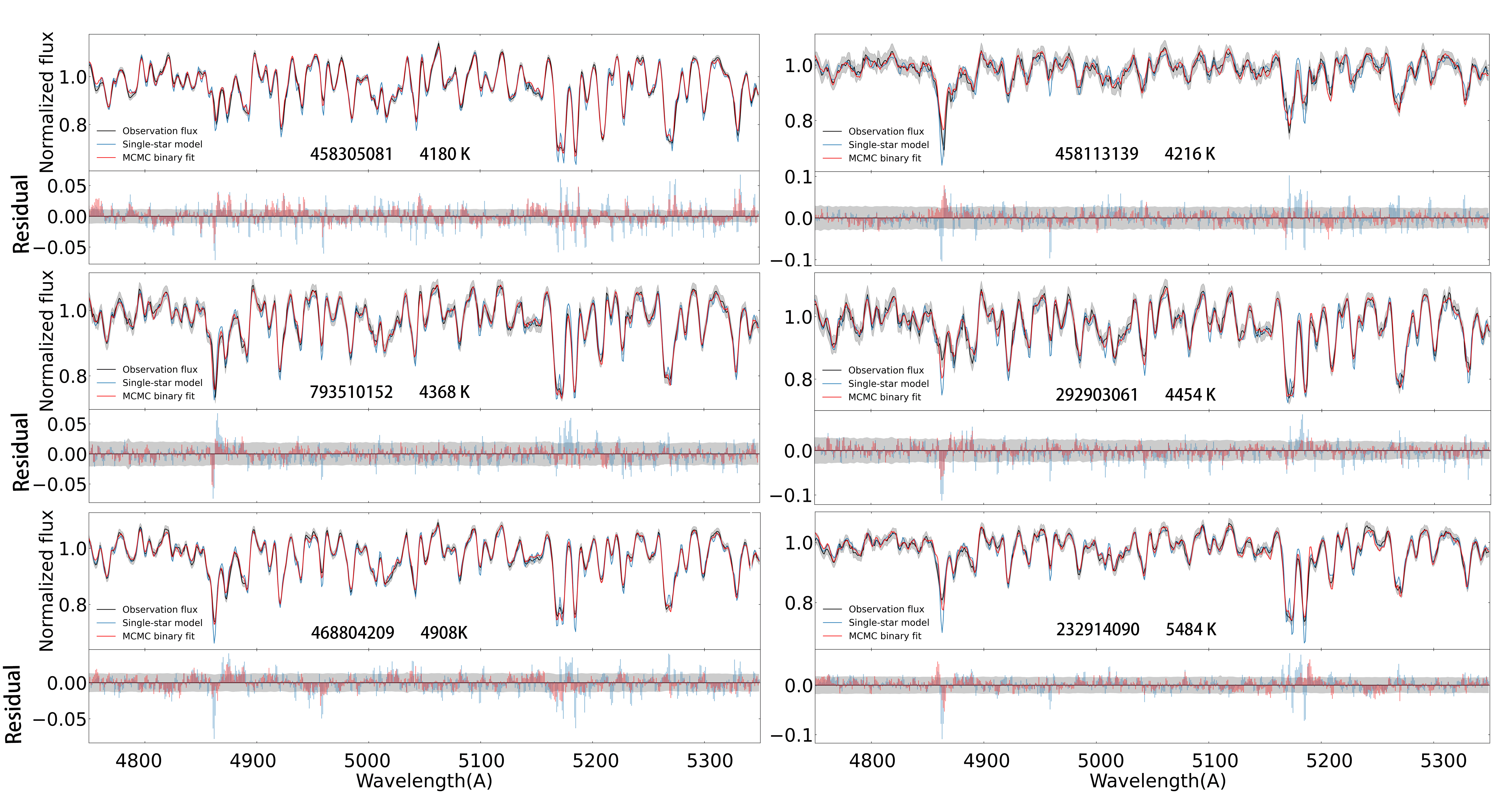}
\caption{Comparison of single-star and binary-star models fitting for binary candidates (with the LAMOST $Obsid$) on the giant branch.}
\label{fig:Gaints_samples}
\end{figure*}

Meanwhile, we found 135 (2.8\% of total binary candidates) binary candidates located on the giant branch among our binary candidates (the rectangle in Figure~\ref{fig:CMD}). As shown in Figure~\ref{fig:Gaints_samples}, we present spectral fitting plots for 6 of them, which also exhibit clear binary features. Therefore, we still retained them in our data release. The reason for their identification is the inclusion of giants in our training sample.

\subsection{Color versus temperature} \label{subsec:color_vs_teff}
\begin{figure*}
\centering
\includegraphics[width = 18.5cm]{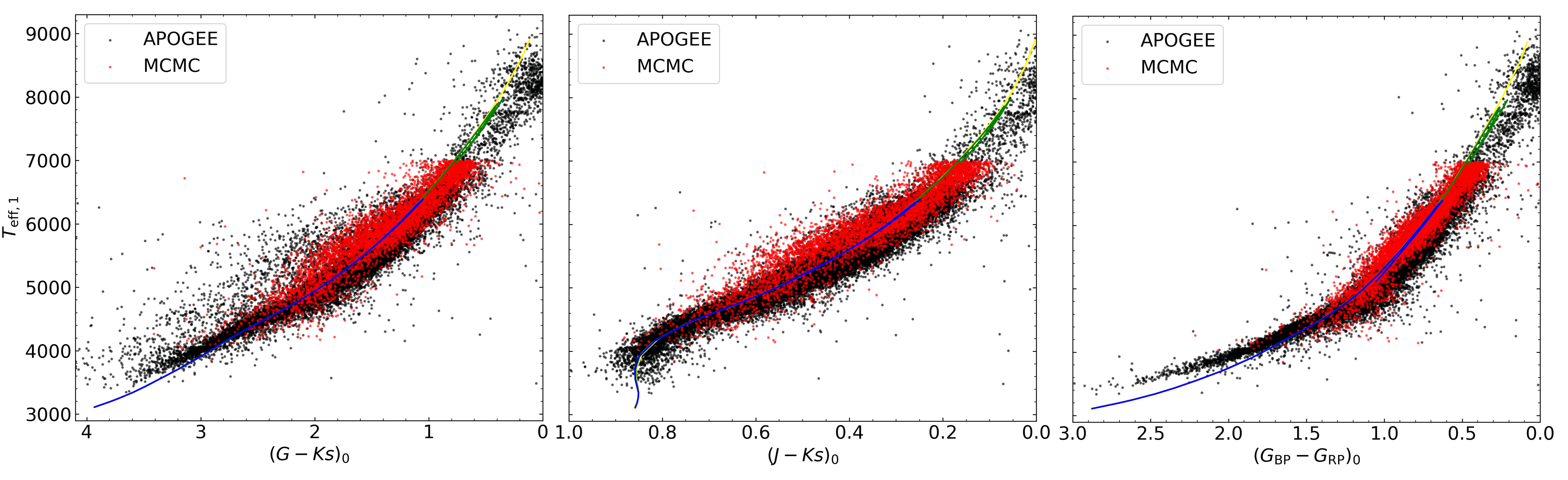}
\caption{The distributions of different colors versus effective temperatures $T_{\rm eff, 1}$ of the primary component. The black points in the background are 10,000 randomly selected field stars from APOGEE. The yellow, green, and blue lines are the isochrone with $\log(\tau)=$ 8.5, 9.0, and 9.5, respectively. The $T_{\rm eff, 1}$ red points are the median values of the posterior distributions in MCMC fitting while the colors are calculated from the photometric observations of $Gaia$~DR3 and 2MASS after extinction correction.}
\label{fig:color_teff}
\end{figure*}

Stellar isochrones, which provide theoretical relationships between color and temperature, offer a way to validate the temperatures obtained through our models. In Figure~\ref{fig:color_teff}, we display the color-temperature relationships in $(G-Ks)_{0}$, $(J-Ks)_{0}$ and $(G_{\rm BP}-G_{\rm RP})_{0}$ for different logarithmic ages $\log\rm \tau$ $= 8.5$ (yellow), 9.0 (green), and 9.5 (blue). These relationships are computed using stellar evolution tracks generated by PARSEC version 1.2S \citep{2012MNRAS.427..127B}. The black points in the background are 10,000 field stars randomly selected from APOGEE, which have cross-matched with $Gaia$ DR3 and 2MASS surveys. As shown in this figure, for these three distributions, the temperature values from our fitting generally agree with those from APOGEE, as well as the theoretical curves, which indicate that our temperature is reliable. 
Furthermore, it's worth noting that at temperatures below 4200~K, especially for $(BP-RP)_{0}-T_{\rm eff,1}$ distribution, the observed data shows slightly higher temperatures compared to the theoretical isochrones. The reason for this inconsistency is that APOGEE overestimates the effective temperature by $\sim$ 110--140~K at $T_{\rm eff} < 4200$~K \citep{2020ApJ...892...31B}. In this work, there are 21 objects in this temperate range, and one should note that the temperatures of these candidates are overestimated because the training parameters are from APOGEE.

\subsection{Photometric observations versus MCMC results} \label{subsec:Photometric}

\begin{figure*}
\centering
\includegraphics[trim=30 0 2 0, clip, width = 18cm]{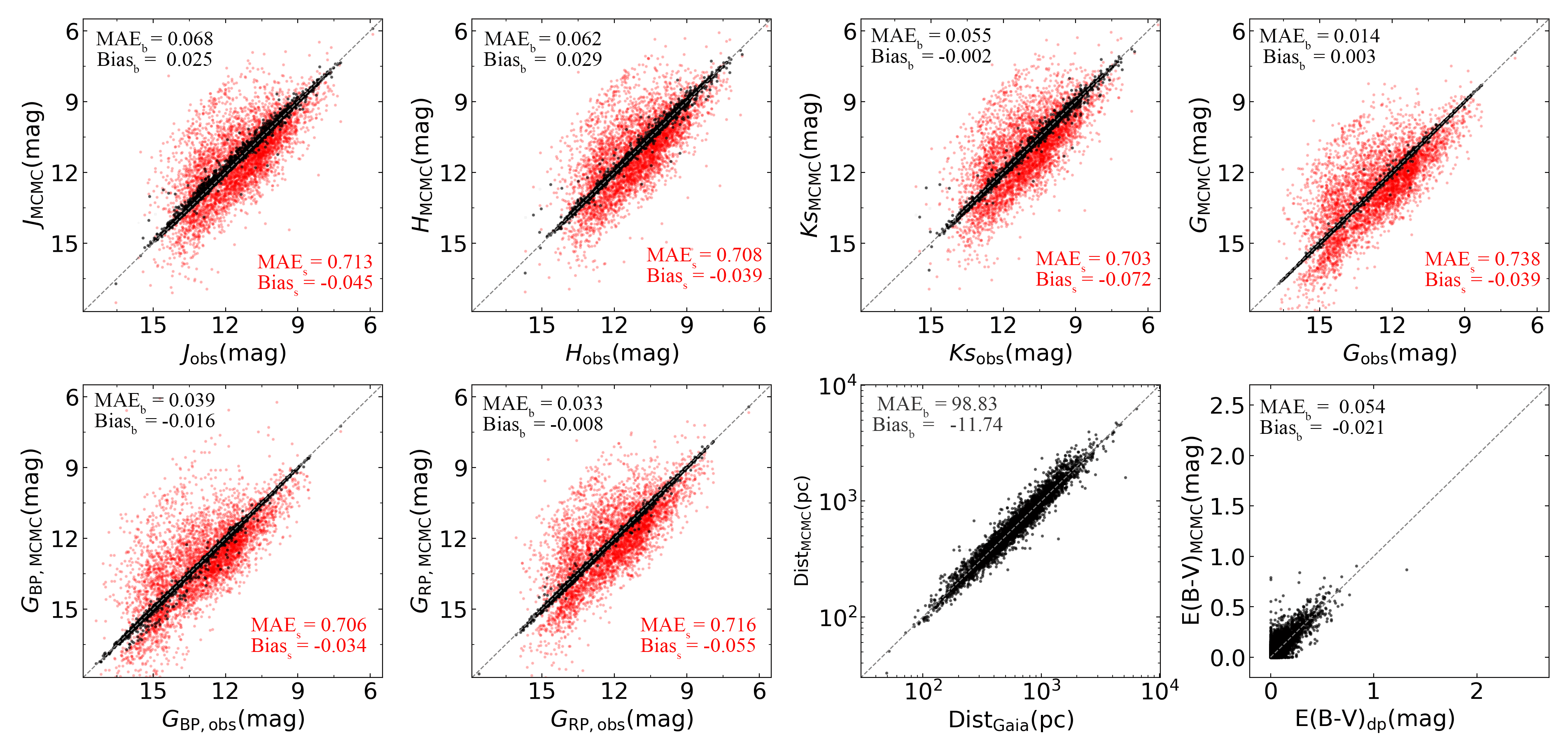} 
\caption{Comparison of photometric, astrometric, and Galactic extinction parameters from observations and the MCMC fitting. The black and red points represent the median values (black points) of the posterior distributions by MCMC and single-star model fitting (red points) results, respectively. The MAE and bias have the same definitions as Figure~\ref{fig:one-to-one_plot}, while the subscripts `b' and `s' represent the MCMC binary fitting and the single-star model fitting, respectively.}
\label{fig:Different_bands_ebv_dis}
\end{figure*}

As described in Section~\ref{subsec:MCMC}, we utilized the photometric,  astrometric, and Galactic extinction information to perform the MCMC fitting jointly. In detail, the photometric observations are from 2MASS ($J$, $H$, and $Ks$ bands) and $Gaia$ DR3 ($G$, $G_{\rm BP}$, and $G_{\rm RP}$ bands), while the distance and $E(B-V)$ are calculated from the parallax of $Gaia$ DR3 and $\tt dustmaps$, respectively. Comparing the median values of each parameter in MCMC fitting with the observed values is a good test of our binary searching results, shown in Figure~\ref{fig:Different_bands_ebv_dis}. Without considering the binary evolution, the luminosity of an SB2 system with two equally luminous component stars is $\sim$0.75~mag higher than that of a single star in the observation. In the first six panels, both $\rm MAE_{\rm b}$ and $\rm bias_{\rm b}$ are significantly smaller than $\rm MAE_{\rm s}$ and $\rm bias_{\rm s}$, which means that the magnitudes of the binary fits are closer to the observed values and have less scatter. Here the subscript `b' represents the binary-model fitting, while the `s' represents the single-star model fitting.
Moreover, the differences between the one-to-one line and $\rm MAE_{\rm s}$ are close to $\sim$0.7 mag in all bands, suggesting that most of the photometric magnitudes derived from the single-star model are fainter than the observed one, which requires the contribution of a secondary star with a large $q$ to match the observed magnitudes. 
For the fitting of distance and $E(\rm B-V)$, the priors are in the non-information normal (flat) distribution with range (0, 100000)~pc and (0, 10)~mag, the values of their $\rm MAE$ and $\rm bias$ indicate that they also follow the one-to-one line tightly with small scatter. In summary, these comparisons mentioned are robust tests for the identification of spectroscopic binary candidates, while the parameters inferred from our MCMC model fitting are reasonably accurate.

\subsection{Orbital parameters} \label{subsec:Joker}

We employed the $\tt Joker$ \citep{2017ApJ...837...20P} package to provide the orbital solutions for part of the identified binary systems. The $\tt Joker$ is a specially created MCMC sampler that identifies circular or eccentric orbits and performs well with nonuniform data. The version of $\tt Joker$ is the branch adapted for SB2s. \footnote{https://github.com/adrn/thejoker@sb2} The input parameters are RVs with corresponding uncertainties (from MCMC results) and the Barycentric Julian Date (BJD; transformed from Modified Julian Day in LAMOST LRS DR9 by $\tt LASPEC$).
We calculated the dynamical parameters for the 56 objects with the number of observations $N_{obs}\ge7$. This choice is based on \citet{2018MNRAS.476..528E} who concluded that the success rate of radial velocity-restricted orbits is greater than 80\% when $N_{obs}\ge7$ and the phase coverage $U_{N}V_{N}>50\%$. The $\tt Joker$ provides well-fitted orbital solutions for 44 of the 56 SB2 candidates (78.57\%), which are listed in Table~\ref{table:Orbital_parameters}, including parameters of $N_{\rm obs}$, spectroscopic period $P_{\rm spec}$, eccentricity $e$, argument of periastron $\omega$, center-of-mass velocity $\gamma$, and the radial velocity semi-amplitudes of the primary and secondary components $K1$ and $K2$.

\begin{table*}
   \caption{Orbital solutions for SB2s with observational epochs $N_{obs}\ge7$}
     \label{table:Orbital_parameters}
   \begin{center}
   \begin{tabular}{rrrrrrrrr}\hline \hline
 R.A. & DEC. & $N_{\rm obs}$ & $P_{\rm spec}$ & $e$ & $\omega$  & $\gamma$ & $K1$ & $K2$ \\
 (degrees) &(degrees) &    & (days) & & (radians) &(km~s$^{-1}$) &(km~s$^{-1}$)&(km~s$^{-1}$)\\
\hline
 59.169521&26.055595&13&11.04081&0.0031&6.002&27.04&-69.09&-74.66\\
101.443596&23.440002&9&1.27834&0.0598&2.067&53.26&87.34&87.50\\
94.471954&22.568149&8&3.86984&0.7827&6.123 &37.95&490.81&510.05\\
57.968642&23.093190&15&0.47818&0.0852&4.071 &58.11&-112.63&-115.97\\
130.704190&15.393808&7&0.28375&0.0120&4.703 &31.89&-170.33&-172.34\\
164.444078&9.978066&12&0.31215&0.0212&1.596&32.28&158.62&158.15\\
161.475796&12.468433&7&9.04761&0.0799&5.005 &4.93&-106.65&-108.99\\
101.859983&22.575474&16&1.21782&0.0067&2.908&-26.57&125.91&137.00\\
102.506908&22.357705&16&0.64418&0.0093&5.590 &79.80&-123.95&-129.92\\
102.572538&22.506170&13&2.70799&0.0942&5.461 &66.03&238.37&238.21\\
...&...&...&...&...&...&...&...&...\\
\hline\noalign{\smallskip}
  \end{tabular}
    \begin{tablenotes}
  \item[1] This table is available in its entirety (with orbital solutions for 44 systems) in machine-readable form.
  \end{tablenotes}
  \end{center}
\end{table*}

\begin{figure}
\centering
\includegraphics[trim=0 0 0 0, clip, width = 8.9cm]{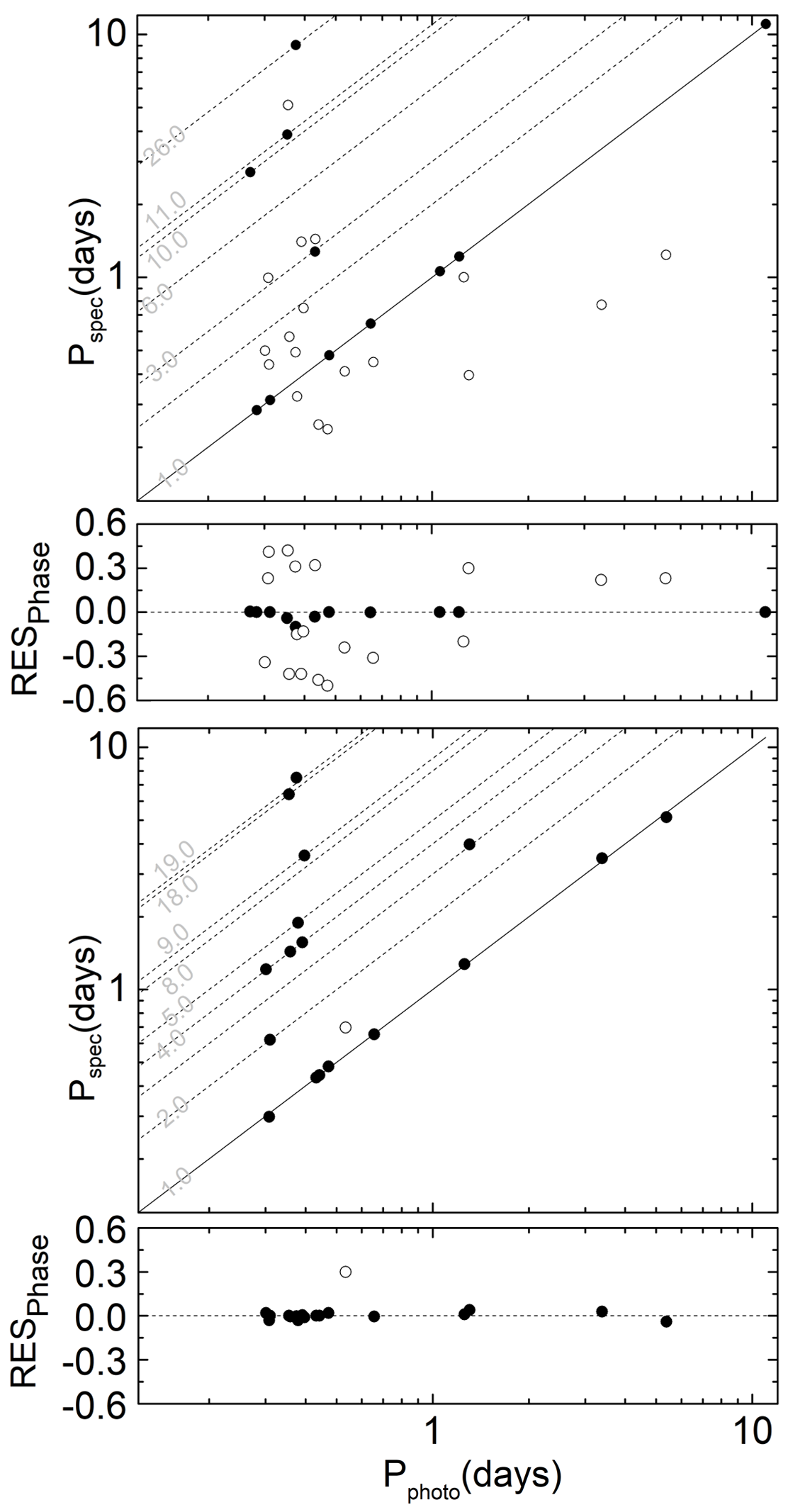}
\caption{Comparison of the $P_{\rm photo}$ collected from ASAS-SN (photometry) and $P_{\rm spec}$ from the $\tt Joker$ (spectrometry). Panels (1) and (3): The solid and dashed lines represent one-to-one and different slope lines, respectively. The number ($N$) indicates the slope of the line, i.e., $P_{\rm spec} = N P_{\rm photo}$, $N=1.0,2.0,3.0,...,26.0$. Panels (2) and (4): The corresponding residuals of panels (1) and (3) were converted to phase form. The points are all binary candidates with $\geq$7 epochs and photometric periods. The solid black points are those with $P_{\rm spec}-N P_{\rm photo} < 0.1P_{photo}$, indicating the spectroscopic period matches well with the photometric one, while the empty circles do not. Panel (1) displays the period distribution results for 29 objects with photometric observations. Panel (3) shows the same distribution, while for 18 objects represented by empty circles in panel (1), $P_{\rm spec}$ were refitted by further reversing the RVs of up to four observations.}
\label{fig:period_compare}
\end{figure}

The period of the binary star can be derived through photometric light curves or kinetic RV curves. These two methods can corroborate each other. In Figure~\ref{fig:period_compare} panel (1), 29 objects with the photometric period $P_{\rm photo}$ from the ASAS-SN are plotted in circles to compare with the spectroscopic period $P_{\rm spec}$ from the $\tt Joker$ fitting. We have labeled 11 sources with black solid circles, accounting for approximately 37.9\% of the total sample, whose $P_{\rm spec}$ is nearly identical to the $P_{\rm photo}$ or an integer multiple of it. 37.9\% is also the percentage of objects with all RVs correctly assigned to component stars after the RV reversal process.
For the other 18 sources represented by the empty circles in panels (1) and (2), we found that by further swapping up to four RVs (in the $\tt Joker$ fitting) based on the sequence of $\delta \chi^{2}_{\rm b,e}$ (in descending order), the $P_{\rm spec}$ of 17 sources can match well with $P_{\rm photo}$ as shown in panels (3) and (4). Only one source remains non-matched periods, likely due to low SNR spectroscopic observations. As described in Section~\ref{subsec:reverse_RV}, it is challenging to assign the radial velocities of binary systems with $q$ close to 1 to the correct component stars completely, because of the low spectral resolution of the LAMOST LRS data, the limited number of spectroscopic observations or the low SNRs.
Nevertheless, 96.55\% (28/29) of the sources exhibit consistent $P_{\rm spec}$ and $P_{\rm photo}$ values, indicating the accuracy of our RV measurements.

\begin{figure}
\centering
\includegraphics[width = 8.7cm]{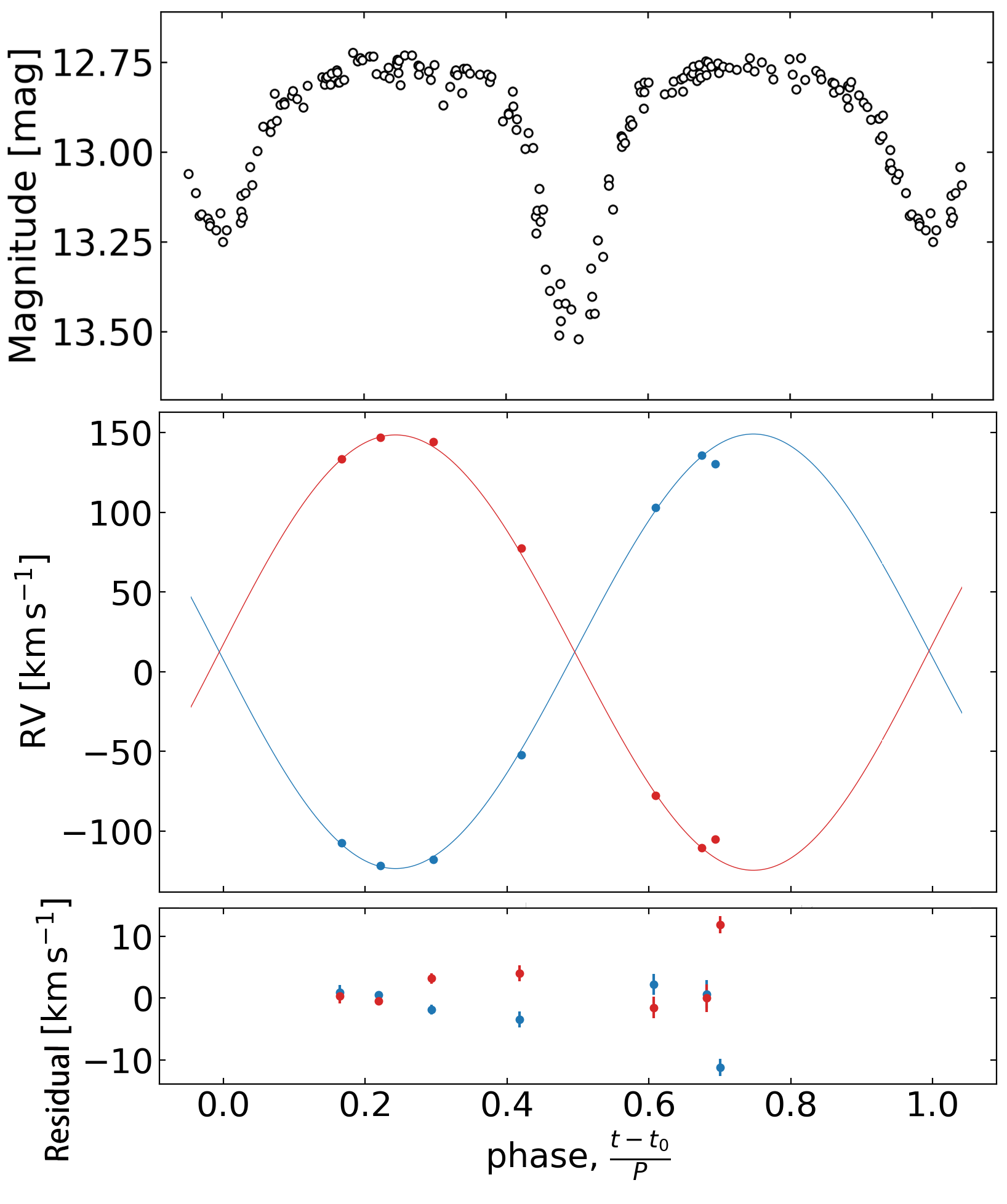}
\caption{The orbital solution of sources J0913. The source serves as an example for the well-constrained system in the orbital fitting. The black points in the upper panel are the photometric light curve from AVSD. The red and blue points represent the phase-folded radial velocities of primary and secondary components.}
\label{fig:light_curve_RV_curve}
\end{figure}

In Figure~\ref{fig:light_curve_RV_curve}, we show an example of the identified binary system with the orbital solution: J0913 with $q=0.98$ and $N_{\rm obs}=7$. The photometric light curves are obtained from the ASAS-SN Variable Stars Database (AVSD) \footnote{https://asas-sn.osu.edu/variables}.

\section{Comparison with Other Catalogs} \label{subsec:ComparisonOtherCatalogs}

\subsection{Comparison with ASAS-SN} \label{subsubsec:ASAS-SN}

\begin{figure*}
\centering
\includegraphics[trim=30 0 20 0, clip, width = 18cm]{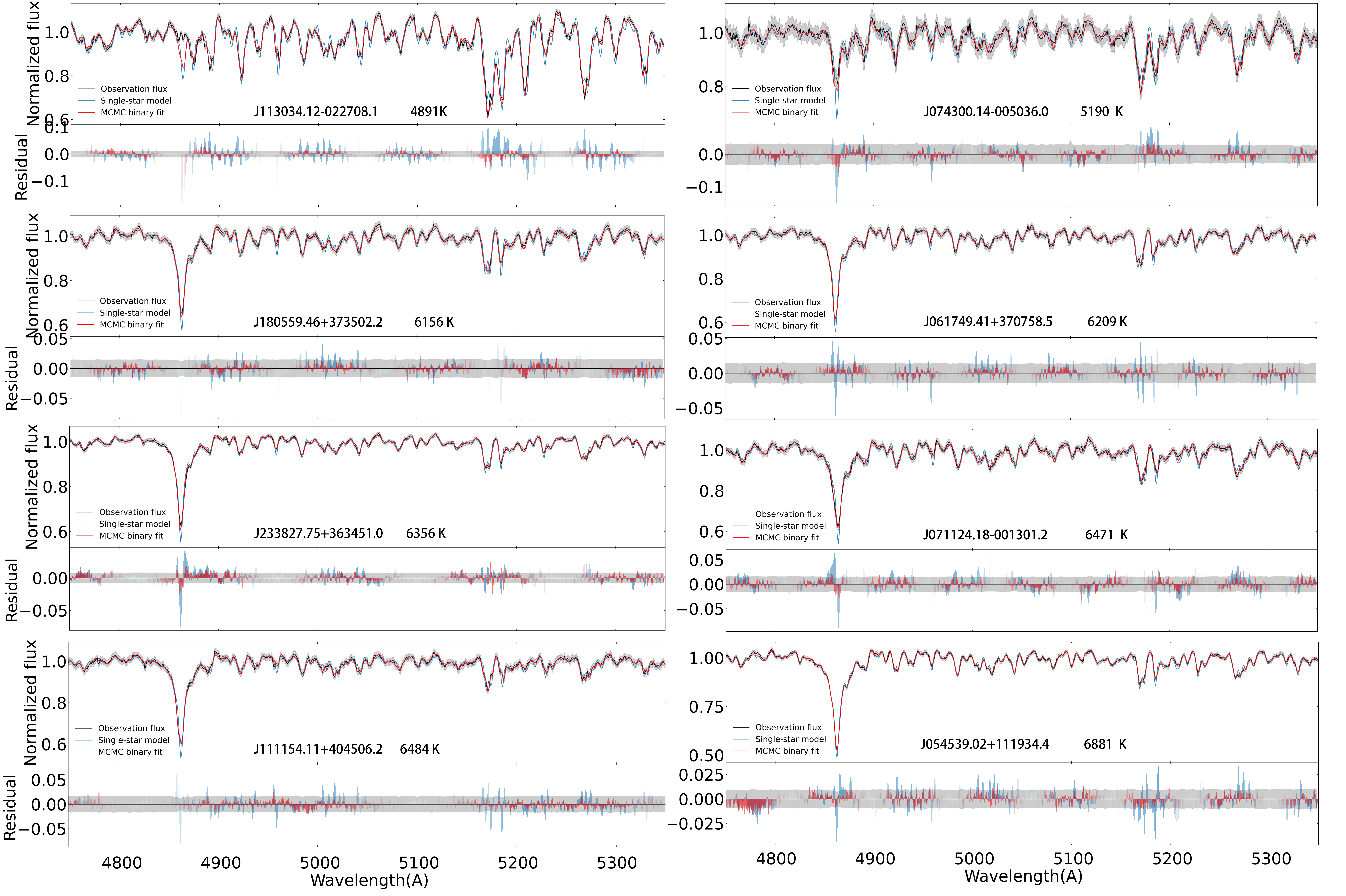}
\caption{Comparison of single-star and binary-star model fittings for the non-occulting binary systems in AVSD.}
\label{fig:ASAS_othertype}
\end{figure*}

Binary stars can also be distinguished from single stars by their eclipsing light curves. ASAS-SN as one of the largest time domain surveys periodically scans the entire visible sky with a cadence of $\sim$2--3 days and a sensitivity limit of $V$ $\lesssim$ 17 mag \citep{2018MNRAS.477.3145J, 2021MNRAS.503..200J}, which provides 687,695 light curves and their corresponding variable star types on the AVSD. In total, 2596 binary star candidates have counterparts in the AVSD, which make up 53.5\% of our SB2 candidate sample. Among these 2596 sources, 1724 objects have only one LAMOST spectroscopic observations, while the other 872 objects have multi-epoch spectroscopy. Their light-curve classifications include $\beta~Persei-$type (Algol) binary (EA), $\beta~Lyrae-$type binary (EB), W Ursae Majoris$-$type binary (EW), ellipsoidal variable star (ELL), rotational variables (ROT), W Virginis type variable with period$<$8~days (CWB), $\delta$ Scuti variables (DSCT), RR Lyrae variables with asymmetric (RRAB) or nearly symmetric (RRC) light curves and variable star of unspecified type (VAR), which are counted in Table~\ref{table:ASAS-SN_classification}. For these classifications, the first four (EA, EB, EW, and ELL) are due to binary occultation, while the other kinds of variables are mainly caused by stellar pulsations or rotation. Among the above variable stars, the percentage of occultation-type variables among all the ASAS-SN counterparts is at least 92.60\%, indicating that most of these counterparts are the true spectroscopic binary stars and our method is reliable in searching SB2s. In Figure~\ref{fig:ASAS_othertype}, we plotted ten spectra classified by ASAS-SN as pulsating or rotating variable stars. For these objects, we found that most of them also have distinctive binary spectral features, which means that they could be SB2s that are misclassified in ASVD. 

Cross-matching the model-fitted 2,619,768 unique LAMOST sources (Section~\ref{sec:preprocessing}) with the 154,600 occulting binaries classified by ASAS-SN (EA, EB, EW, and ELL types) yields a total of 13,051 counterparts. So our method identified 19.9\% (2596/13051) of them. The possible reasons for these binaries that are covered by LAMOST observations but not identified by our spectral models may include: 1) relatively small-mass secondary star with insignificant contributions to the spectra; 2) the maximum RV offset between binary components is less than 100 km$\,$s$^{-1}$; 3) the binary is in phase near superior or inferior conjunction and the RV offset is small; 4) the inclination of the binary system is small.

\subsection{Comparison with TESS} \label{subsubsec:TESS_catalog}

The Transiting Exoplanet Survey Satellite (TESS; \citealt{2014SPIE.9143E..20R}) is a space-based survey mission designed to search for exoplanets, while also serves other fields of time-domain astronomy such as supernovae, active galactic nuclei, and variable stars. Binary star identifications have been carried out using TESS light curves (e.g., \citealt{2022ApJS..258...16P,2023MNRAS.522...29G}).

\citet{2022ApJS..258...16P} provided a sample of 4,584 binary systems selected from the initial two years of TESS observations. To categorize the light curve profiles, they utilized a parameter ``morphological coefficient", originally introduced by \citet{2012AJ....143..123M}. This coefficient spans a continuous range of values, ranging from approximately 0 (indicating widely separated stars with minimal ellipsoidal modulation) to 1 (suggesting either contact binaries or purely ellipsoidal light curves). Following a cross-matching process with their dataset, 88 counterparts were identified.

\citet{2023MNRAS.522...29G} presented 15,779 binary candidates that consist of main-sequence primary stars and have orbital periods under 5 days. They identified these binary candidates using TESS full-frame image light curves based on the tidally induced ellipsoidal modulation, ensuring homogeneity across the sample. After cross-matching, we obtained 335 counterparts.

There are 15 duplicates in the above two cross-matches, so we have 408 binary candidates matching with those (\citealt{2022ApJS..258...16P,2023MNRAS.522...29G}) found based on TESS optical light curves. This represents 8.42\% of our sample size, which is approximately twice as much as the result of $\sim$4.6\% of \citet{2021AJ....162..184K}. They identified 8,105 multiple-star system candidates in APOGEE based on the CCF method, out of which 369 have apparent binary eclipsing features in the whole TESS footprint. 

 \begin{table*}
   \caption{Object number in ASAS-SN light-curve classifications of binary star candidates.}
     \label{table:ASAS-SN_classification}
   \begin{center}
   \begin{tabular}{cccccccccc}\hline \hline
Classification      & EA/EB/EW/ELL & ROT & CWB & DSCT &HADS & RRAB & RRC & VAR & SUM \\
Single observations &1591          & 66  & 3   & 2    & 1   & 2     & 9  &  50 & 1724\\
Multiple observations & 813        & 27  &---  & --   & 1   &---   & 5   &  26 & 872\\
\hline\noalign{\smallskip}
  \end{tabular}
  \end{center}
\end{table*}
\subsection{Comparison with LAMOST Medium-Resolution Survey} \label{subsec:LMR_catalogues}

Based on the LAMOST Medium Resolution survey (MRS), there have been several efforts to search SB2s using machine learning, such as \citet{2022ApJS..258...26Z}. They developed a convolutional neural network model trained on simulated spectra of single and binary systems, synthesized using the MIST stellar evolution models and ATLAS9 atmospheric models, to identify SB2s. They released a catalog of binary probabilities for more than 1 million sources. Their model achieves a novel theoretical false-positive rate by incorporating an appropriate penalty with the penalty parameters $\Lambda$ on negative samples. They used $p_{q}^{\Lambda}$ to denote the probability (from 0 to 1) of a source being an SB2 at the penalty factor $\Lambda$ and the $q$-th percentile. They released the binary probability of each source at the value of $\Lambda$ = 8, 16, and 32, and $q$ = 0, 16, 50, 84, and 100. By adopting the criteria of $p_{q=0}^{\Lambda=16} > 0.5$ and bad pixel number $npixbad < 100$ for both red and blue arms of spectra as those in \citet{2022ApJS..258...26Z} to select binary candidates, we obtained 231 identical sources within 3\arcsec. 

We found another 198 binary candidates (in \citealt{2022ApJS..258...26Z}'s whole catalog) classified by our binary catalog and ASAS-SN as binary but not in those 231 objects. Then, by relaxing the criterion that $p_{q=100}^{\Lambda=16} > 0.5$ in either arm (instead of in both arms), we obtained 455 identical objects with \citet{2022ApJS..258...26Z}. This proves that the number of binary candidates 2318 given by \citet{2022ApJS..258...26Z} is a lower limit, and more binary candidates can be obtained by different screening criteria.

\subsection{Comparison with APOGEE} \label{subsubsec:APOGEE}

Cross-matching our binary candidates with APOGEE SB2 candidates from \citet{2018MNRAS.476..528E} and \citet{2021AJ....162..184K} yielded 9 and 58 counterparts, respectively. The main reason for this is caused by the low number of 90915 common sources between APOGEE DR16 (\citealt{2018MNRAS.476..528E} from DR13) and LAMOST LRS DR9. Since our final binary candidates are only identified from sources with $\log g > 3.5$, the number of common sources is left at $\sim$46,000. 
Based on our detection rate (0.14\%), 64 systems (46000$\times$0.14\%) in the APOGEE DR16 sample (and even fewer for APOGEE DR13) are expected to be identified, which is consistent with the number of counterparts (58) in \citet{2021AJ....162..184K}.
 
 Other reasons might prevent us from identifying SB2s from APOGEE. The LAMOST LRS has a much lower spectral resolution than that of APOGEE ($R\sim22,500$). The lower resolution leads us to identify binary candidates with component velocity differences greater than 100~km~s$^{-1}$. The need for large RV offset also means that our results lack long-period (>10 days) sources, which is what \citet{2018MNRAS.476..528E}’s sample specializes in. Furthermore, the NIR spectra of APOGEE have advantages in identifying systems with smaller mass ratios ($q>0.4$ in \citealt{2018MNRAS.476..528E}) compared to optical spectra ($q\geq0.7$ in our work) since the less massive component could have a higher fraction of contribution to the NIR spectra than to the optical spectra.

\subsection{Comparison with $Gaia$ DR3} \label{subsubsec:GaiaDR3}

$Gaia$ DR3 includes the two-body orbital solutions for systems covering photometric, spectroscopic, and transiting binaries, as well as combinations of the three. Out of 443205 systems that they released in the orbital solution catalog ($gaiadr3.nss\_two\_body\_orbit$)\footnote{https://doi.org/10.17876/gaia/dr.3/74}, we have 39 counterparts in our catalog with a cross-matching radius of 3\arcsec. The limited number of matches may be because most binaries ($>70\%$) discovered by $Gaia$ DR3 have longer periods ($>10$ days), which is outside the optimal range for our method.

\section{Summary} \label{Summary}

In this work, we have developed a data-driven machine learning model that utilizes simultaneous fittings of single-star and binary-star models on observed spectra to search for spectroscopic binary stars, applying it to LAMOST LRS DR9. We released two catalogs: one (Table~\ref{table:4848 binary candidates}) containing the physical parameters of 4,848 binary stars, and another (Table~\ref{table:Orbital_parameters}) with the dynamical parameters of 44 sources. Our main results are as follows:

\begin{enumerate}
\item In this work, the physical parameters provided for the identified 4848 binary star candidates include not only atmospheric parameters such as effective temperature (4000~K $\lesssim T_{\rm eff}\lesssim$ 7000~K), surface gravity (3.5 $\lesssim \log g \lesssim $ 5.0), and metallicity (-2.0 $\lesssim$[M/H]$\lesssim$ 0.5) but also parameters like mass, radius, and age. In addition to spectral data, photometry, parallax, and extinction information are simultaneously fitted in the MCMC process. Therefore, we also present the distance and extinction parameters. The photometric magnitudes obtained from the binary-star model with the MIST are more consistent with the observed values than those from the single-star model, showcasing the validity of our SB2 identification method. 
\item Our results are primarily focused on main-sequence binary stars of FGK types (4713 objects). Additionally, there are 135 binary candidates with giant components. This property arises due to the skewed distribution of our training set. Nearly all of these candidates are located above the main sequence in the color-magnitude diagram (Figure~\ref{fig:CMD}), which suggests their binary nature. It's important to note that other types of binaries, such as white dwarf–main sequence binaries and white dwarf–white dwarf binaries, are not investigated in this study.

\item Given the LRS spectral resolution of $R\sim1800$, our model exhibits favorable identification performance for sources with velocity differences exceeding 100 km$\,$s$^{-1}$, as inferred from semi-empirical spectral experiments. This value is lower than the 50 km$\,$s$^{-1}$ velocity resolution of the MRS. Due to the presence of numerous sources with high mass ratios in our sample, the assignment of radial velocities to incorrect member stars unavoidably occurs. In this study, we attempt to mitigate this problem by using a method of radial velocity reversal to obtain the minimum $\chi^{2}_{b,e}$ value for the spectral fit. We provide $\delta \chi^{2}_{\rm b,e}$ values before and after the radial velocity inversion in the catalog to assess its impact on spectral fitting.
 Our model is well-suited for cases with mass ratios $q\geq0.7$. The completeness of the sample increases with the rise of the mass ratio, with a slight decline as it approaches 1.
\item We provide orbital solutions for 44 short-period ($\lesssim 10\,$days) objects (Table~\ref{table:Orbital_parameters}) with sufficient ($\geq7$) observational epochs. Without performing additional radial velocity swap, approximately 37.9\% of the results are consistent with integer multiples of the periods provided by ASAS-SN (upper panel of Figure~\ref{fig:period_compare}). After performing further RV swap operations in radial velocity curves fitting, the majority of the sources' radial velocity period solutions can align with integer multiples of the photometric periods (lower panel of Figure~\ref{fig:period_compare}). 

\end{enumerate}

The model we obtained can be applied to subsequent LAMOST low-resolution data releases with larger sample sizes to acquire more spectroscopic binary star samples. The increase in sample size is of significant importance for understanding the properties of SB2s, such as the characteristics of stellar multiplicity and the co-evolution of binary stars. For some binary star candidates, we will conduct follow-up observations to further analyze their atmospheric and orbital parameters. Additionally, expanding the methods applicable to the binary star population in this work involves extending the training samples of the single-star model to both lower and higher temperature ranges. This can be achieved by increasing the number of observed spectral learning samples or by incorporating theoretical spectra.

\acknowledgements

We are grateful to the anonymous referee for providing thoughtful and helpful comments that improved the manuscript. This work is supported by the National Natural Science Foundation of China (NSFC) under grant numbers 12273029 and 12221003 (J.L. and J.W.). B.Z. acknowledges the support from Natural Science Foundation of China (NSFC) under grant No.12203068.
We acknowledge the data support from the Guoshoujing Telescope. Guoshoujing Telescope (the Large Sky Area Multi-Object Fiber Spectroscopic Telescope LAMOST) is a National Major Scientific Project built by the Chinese Academy of Sciences. Funding for the project has been provided by the National Development and Reform Commission. LAMOST is operated and managed by the National Astronomical Observatories, Chinese Academy of Sciences.

\software{LASPEC\citep{2020ApJS..246....9Z, 2020RAA....20...51Z}, Matplotlib\citep{Hunter:2007}, joblib\citep{joblib_}, NumPy\citep{harris2020array}, SciPy\citep{2020SciPy-NMeth}, Astropy\citep{2013A&A...558A..33A, 2018AJ....156..123A, 2022ApJ...935..167A}}

\bibliography{reference}{}
\bibliographystyle{aasjournal}

\end{document}